\documentclass[superscriptaddress,longbibliography, reprint,amsmath,amssymb,aps,prl,floatfix]{revtex4-2}
\usepackage[margin=1.5cm]{geometry}

\usepackage{xcolor} 
\usepackage{graphicx}
\usepackage{dcolumn}
\usepackage{upgreek}
\usepackage{bm}
\usepackage{hyperref}
\usepackage{floatrow}
\usepackage{times}
\usepackage[mathlines]{lineno}
\usepackage{physics}
\usepackage{color,soul}

\usepackage{etoolbox}
\usepackage{comment}

\makeatletter

\newcommand{\cw}[1]{\textcolor{blue}{#1}}

\definecolor{custompurple}{RGB}{112, 48, 160}
\definecolor{customblue}{RGB}{21, 96, 130}

\makeatother

\begin{document}

\title{Non-Markovian Relaxation Spectroscopy of 
Fluxonium Qubits}
\author{Zetong Zhuang}
\author{Dario Rosenstock}
\email{Present address: Google Quantum AI, Santa Barbara, CA, USA}
\author{Bao-Jie Liu}
\affiliation{Department of Physics, University of Massachusetts-Amherst, Amherst, MA, USA}
\author{Aaron Somoroff}
\email{Present address: SEEQC, Inc., Elmsford, New York, USA}
\affiliation{Department of Physics, University of Maryland, College Park, MD, USA}
\author{Vladimir E.~Manucharyan}
\affiliation{Institute of Physics, Ecole Polytechnique Federale de Lausanne, Lausanne, Switzerland}
\author{Chen Wang}
\email{Correspondance should be addressed to: wangc@umass.edu}
\affiliation{Department of Physics, University of Massachusetts-Amherst, Amherst, MA, USA}

\date{\today}

\begin{abstract}

Recent studies have shown that parasitic two-level systems (TLS) in superconducting qubits, which are a leading source of decoherence, can have relaxation times longer than the qubits themselves.  However, the standard techniques used to characterize qubit relaxation 
is only valid for measuring $T_1$ under Markovian assumptions and could mask such non-Markovian behavior of the environment in practice.  Here, we introduce two-timescale relaxometry, a technique to probe the qubit and environment relaxation simultaneously and efficiently. 
We apply it to high-coherence fluxonium qubits over a frequency range of 0.1-0.4 GHz, and reveal a discrete spectrum of TLS with millisecond lifetimes. 
Our analysis of the spectrum is consistent with a random distribution of TLS in the aluminum oxide tunnel barrier of the Josephson junction chain of the fluxonium with a spectral and volume density and average electric dipole similar to previous TLS studies at much higher frequencies.  
Our study suggests that investigating and mitigating TLS in the junction chain is crucial to the development of various types of noise-protected qubits in circuit QED. 

\end{abstract}

\maketitle

\subsection{Introduction}

Decoherence of physical qubits in most quantum technology platforms is usually modeled via interactions with a Markovian environment bath.  A Markovian bath relaxes to a fixed equilibrium on short timescales, and hence has no memory of its prior interaction with the qubits.  The widely adapted concepts and measurement protocols for characterizing qubit decoherence, such as the $T_1$ relaxation times, the $T_2$ pure-dephasing times, and dephasing noise spectrum, are all based on the Markovian assumption~\cite{krantz_quantum_2019}.  Environmental degrees of freedom with long relaxation or coherence times can lead to unexpected spatio-temporal correlations in the system dynamics and pose challenges to device characterization and to the implementation of quantum algorithms and error correction protocols~\cite{ white_demonstration_2020, white_non-markovian_2022, shrikant_quantum_2023, brillant_randomized_2025}.  
While the effect of non-Markovian noise has been frequently discussed in the theoretical literature and studied in relatively contrived demonstrations (where the environment is engineered to be non-Markovian 
)~\cite{norris_qubit_2016, figueroa-romero_randomized_2021, goswami_experimental_2021, wang_intrinsic_2021, ahn_non-markovian_2023, wudarski_characterizing_2023}, it remains unclear to what extent the effect of non-Markovianity is at play in the leading physical qubits under pursuit today.  

The superconducting fluxonium qubit~\cite{manucharyan_fluxonium_2009} recently emerged as a promising building block for quantum processors.  When operated at the half flux quantum, the fluxonium features much lower frequency and higher anharmonicity than the standard transmon qubits, and has demonstrated among the best coherence times ($\gtrsim1$ ms~\cite{somoroff_millisecond_2023, wang_achieving_2024}) and two-qubit gate fidelities ($\gtrsim$ 99.9\%~\cite{ding_high-fidelity_2023, lin_24_2025, zhang_tunable_2024}) in superconducting circuits.  
On the other hand, the fluxonium presents an intriguing case where non-Markovian effects may play a more central role in its coherence property.  Its more sophisticated device construction (requiring a chain of Josephson junctions~\cite{manucharyan_fluxonium_2009, pop_coherent_2014} or long strips of high-kinetic-inductance materials~\cite{hazard_nanowire_2019, spiecker_two-level_2023}) makes it more likely to host a large number of resonant two-level systems (TLS)~\cite{muller_towards_2019}; 
the lower qubit frequency ($<$1 GHz) means that these resonant TLS may potentially have slower dynamics.  Indeed, a recent study of a fluxonium qubit made of granular aluminum showed striking non-Markovian qubit relaxation dynamics which is attributed to a TLS bath with millisecond relaxation times compared to a relatively short qubit lifetime on the order of 10 $\mu$s~\cite{spiecker_two-level_2023}.  

In order to resolve and to characterize non-Markovian effects in qubit relaxation, the measurement protocol has to incorporate the state of the environment bath as a dynamic variable.  As is demonstrated in \cite{spiecker_two-level_2023} and in recent studies of photonic-bandgap-engineered transmons~\cite{odeh_non-markovian_2025, chen_phonon_2024}, this can be achieved by controlled preparation of the bath to different initial states 
before proceeding with a standard (pump-delay-probe) qubit $T_1$ measurement that would reveal joint relaxation of the qubit and the bath.  
However, such measurements have low data throughput rate 
due to the slow repetition cycle necessary for the bath to start from equilibrium and the long wait times within the qubit pump-and-probe measurements.  In order to more broadly assess potential non-Markovian relaxation dynamics in various qubits, more efficient measurement protocol needs to be developed.  This can be particularly relevant for understanding the coherence properties of low-frequency fluxonium qubits, given their high promise but large variability reported in the literature~\cite{wang_achieving_2024, nguyen_high-coherence_2019}. 
 
In this work, we introduce a measurement technique designed to probe the qubit and bath relaxation simultaneously and efficiently, which we refer to as two-timescale relaxometry.  The key principle is to consistently extract information on the qubit transition (decay or excitation) rates on short measurement intervals while attempting to polarize the bath on long timescales.  This technique requires efficient reset of the qubit but does not require high-fidelity quantum non-demolition (QND) readout, and hence may be adapted to a variety of qubits. 

We apply the two-timescale relaxometry in a spectroscopic study of fluxonium qubits in the frequency range of 125-400 MHz.  
We find that the qubit relaxation dynamics is governed by a discrete spectrum of TLS with millisecond lifetimes (but poor $T_2$) on top of a background Markovian $T_1$ also in the millisecond range. 
With a detailed finite-element modeling of the dielectric participation ratio of the fluxonium mode, we find that the relaxation spectrum is consistent with a random distribution of TLS in the aluminum oxide tunnel barrier of the fluxonium's Josephson junction chain with an average area density of 0.4 GHz$^{-1}\mu$m$^{-2}$ and average effective dipole moment of 6 Debye.  
Our study therefore shows that the TLS in the 100's of MHz frequency range, which would typically affect other quantum circuits as ``thermal fluctuators", have similar density and electric dipoles as the more widely investigated TLS at multi-GHz frequencies~\cite{muller_towards_2019} and indeed have slow switching dynamics. 
We conclude by discussing implications of these results on future advance of coherence times of fluxoniums and other protected superconducting qubits.

\begin{figure}
    \centering
    \includegraphics[width=1\linewidth]{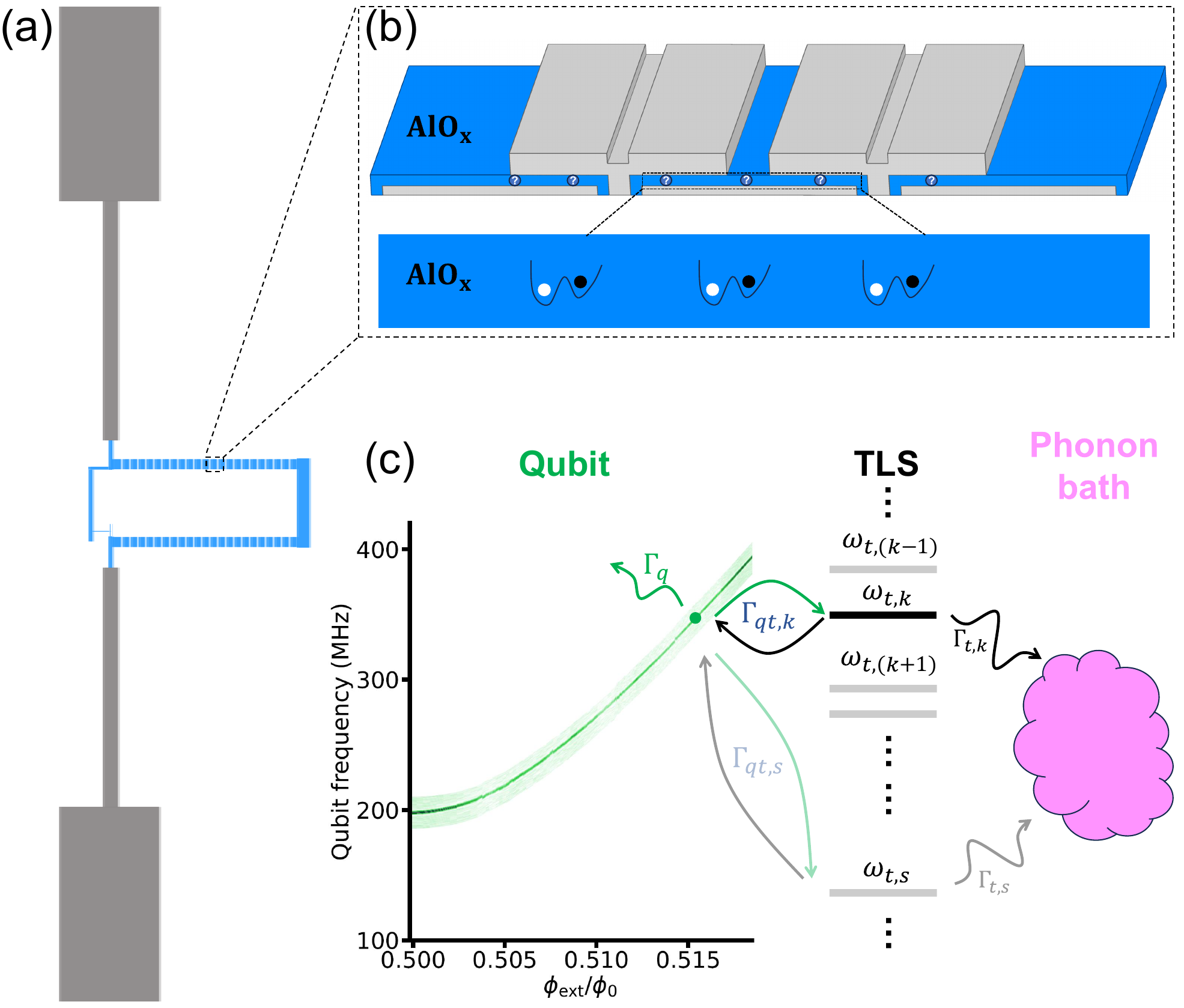}
    \caption{\textbf{TLS in the fluxonium junction chain.}
    \textbf{(a)} Schematic of a fluxonium qubit.
    \textbf{(b)} Schematic of TLS residing in oxide layers of fluxonium’s junction chain.
    \textbf{(c)} When parking fluxonium at a certain frequency ($\omega_q$) through tuning external flux $\phi_{ext}$, its energy relaxation rate will be greatly affected by one or more TLS in junction chain, denoted as $\Gamma_{qt,k}$. All TLS have their own intrinsic decay rate $\Gamma_t$ to their phonon bath. Other environmental factors, as well as the TLS bath that are weakly coupled to the fluxonium qubit at that frequency, contribute to $\Gamma_q$, which is considered the qubit's background decay rate. 
	}\label{fig:dev_schematic}
\end{figure}

\subsection{Device and system}
Our main study is carried out on a fluxonium qubit (Fig.~\ref{fig:dev_schematic}(a)) inside a 3D copper cavity, which has shown coherence times that exceed 1 ms at the half-flux point in a previous study~\cite{somoroff_millisecond_2023}.  The fluxonium has $E_J/2\pi=4.88 $ GHz, $E_C/2\pi=1.09$ GHz, $E_L/2\pi=0.56$ GHz, half-flux frequency ${\omega_{01}} /2\pi= 198$ MHz, and features a 166 Al/AlO$_x$ Josephson junction chain. 
The best coherence times observed in this study are slightly lower than in Ref.~\cite{somoroff_millisecond_2023} due to stochastic presence of TLS near the half flux point and less rigorous filtering and shielding in a different cryogenic setup in a different lab, but nonetheless reaches $T_{2E}$ above 0.6 ms.  
We also measured another fluxonium device in a planar package with half-flux $T_{2E}$ times in the range of 70 to 90 $\mathrm{\mu s}$, whose details are discussed in Supplementary Section I. 

\begin{figure}[tbp]
    \centering
    \includegraphics[width=1\linewidth]{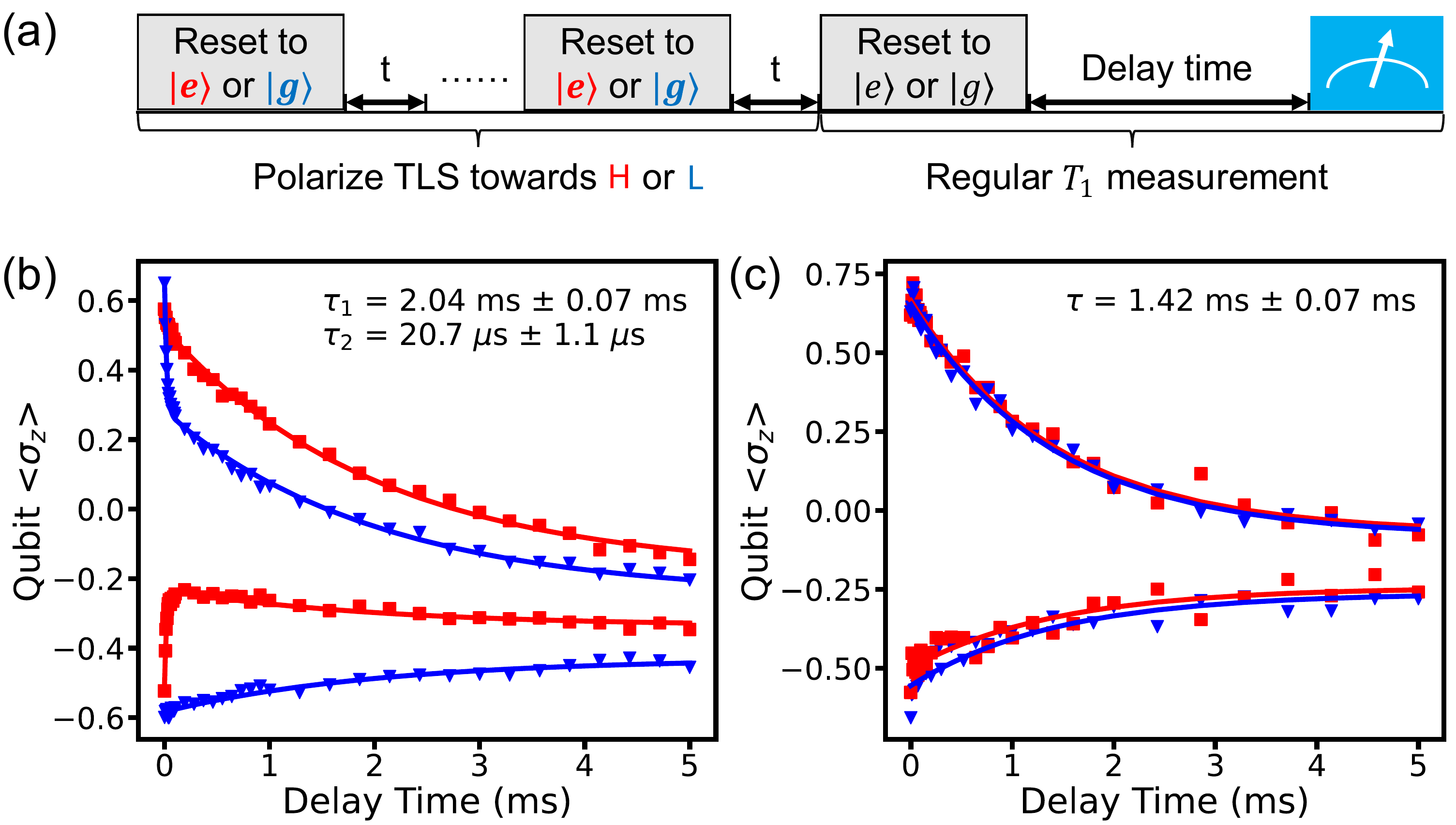}
    \caption{\textbf{Qubit $T_1$ measurement following different bath-preparation sequence.}
    \textbf{(a)} Pulse sequence diagram. Before regular $T_1$ measurement, we polarize TLS state by repeatedly initializing qubit to $\ket{e}$ or $\ket{g}$ state and allowing for a qubit-TLS energy exchange time $t$. 
    In this example, we use $t=25\mu$s and reset the qubit 7 times. 
    After the TLS bath is prepared, regular $T_1$ measurements are done with qubit initialized in $\ket{e}$ or $\ket{g}$ states.
    \textbf{(b)} Red dataset is taken as TLS is polarized towards a high-energy state (H) 
    and blue one towards a low-energy state (L). 
    Each trace is fit with the sum of exponential decays using the same two time scales. One is qubit-TLS interaction rate, i.e. $\Gamma_{qt}$. The other is the qubit and TLS decaying together, i.e. $\frac{\Gamma_{q} + \Gamma_{t}}{2}$. 
    \textbf{(c)} Same experimental process is done at different flux point. As red and blue traces overlap, it shows that there the environment is not polarizable 
    (over the timescale of the given protocol) at this qubit frequency.
	}\label{fig:conditional_T1}
\end{figure}

The large volume of aluminum oxide in fluxonium's junction chain potentially provides a rich soil for discrete TLS with significant interaction to the qubit (Fig.~\ref{fig:dev_schematic}(b)). 
While this interaction is not expected to reach the strong coupling condition $g_k > \Gamma_{2k}$, (where $g_k$ is the flip-flop interaction between the qubit and the $k$-th TLS, and $\Gamma_{2k}$ is their combined decoherence rates,) it can lead to an incoherent energy exchange rate $\Gamma_{qt,k} = \frac{2g_k^2\Gamma_{2k}}{{\Gamma_{2k}}^2+\Delta_k^2}$ that dominates over other relaxation timescales of the system.  Here  $\Delta_k$ is frequency detuning between qubit and the TLS.  
Generally, in this incoherent limit, qubit-TLS relaxation dynamics can be described by the Solomon equations~\cite{solomon_relaxation_1955, spiecker_solomon_2024}:
\begin{align}
\frac{dZ}{dt} &= -\Gamma_q (Z - Z^{eq}) - \sum_{k}{\Gamma_{qt,k}(Z - p_k)} \\
\frac{dp_k}{dt} &= -\Gamma_{t,k} (p_k - p_k^{eq}) - \Gamma_{qt,k}(p_k - Z)
\label{eq:solomon}
\end{align}
where $Z=\langle\sigma_z\rangle$ represents the qubit polarization in energy eigenbasis, with $Z^{eq} = \langle \sigma_z\rangle^{eq} $ its equilibrium value at long times. $p_k$ is the polarization of the $k$-th TLS, with $p_k^{eq}$ its corresponding equilibrium polarization. $\Gamma_q$ is the ``background" decay rate of the qubit, which results from the qubit interacting with a continuum bath of TLS that are not explicitly modeled and other noise sources such as quasiparticles and radiation modes, and $\Gamma_{t,k}$ are the decay rates of the TLS to their relaxation channels, e.g.~the phonon bath. 
A general solution to these equations provides the qubit with non-Markovian relaxation dynamics, which could manifest as multiple exponential relaxation timescales in experiments if the entire coupled system can be reproducibly initialized.

\subsection{Warm-up example of non-Markovian qubit relaxation}
To illustrate a prominent case of non-Markovian relaxation dynamics, we show qubit $T_1$ relaxation measurements following different bath-preparation sequences. As shown in Fig.~\ref{fig:conditional_T1}(a), before we perform regular qubit $T_1$ measurements, we repeatedly reset the qubit to the (nominal) $\ket{e}$ or $\ket{g}$ states and let it interact with the environment bath for a small time period $t$, similar to Refs.~\cite{spiecker_two-level_2023, odeh_non-markovian_2025}.  Here, the qubit reset is realized with an ad hoc strategy of applying a non-QND readout pulse in the cavity~\cite{bista_readout-induced_2025}, giving reset fidelity in the range of 60\% to 95\% (mean reset fidelity 72\%) (see Supplementary section VI for details). 
Such bath-preparation sequences can saturate the TLS bath towards either their ground or excited state, and the subsequent measurement of the qubit decay is carried out with the qubit prepared in either $\ket{g}$ or $\ket{e}$.  When the bath contains a prominent long-lived TLS (Fig.~\ref{fig:conditional_T1}(b)), we can use a double-exponential model to capture all four decay curves, where the fast time scale represents the qubit-TLS interaction rate and the slow time scale corresponds to their average intrinsic decay rate. Such non-Markovian feature can be easily overlooked and lead to misinterpretation of the qubit relaxation dynamics if traditional $T_1$ protocols are used without attention to the state of the bath (see Supplementary Section II for examples). In comparison, at a different flux-bias point, the bath-preparation sequences may hardly affect the qubit relaxation curve (Fig.~\ref{fig:conditional_T1}(c)), indicating no appreciable memory of the bath.  

Although this method illustrates the non-Markovian relaxation dynamics clearly, it is very slow: It requires a relatively long bath-preparation sequence without acquiring data, and the average interval of measurement is on the order of the slowest time scale of system, which is milliseconds in our device and generally is not known a priori.  This limitation makes it difficult to characterize the TLS environment for a wide frequency range before it reconfigures. 

\begin{figure}[tbp]
    \centering
    \includegraphics[width=0.95\linewidth]{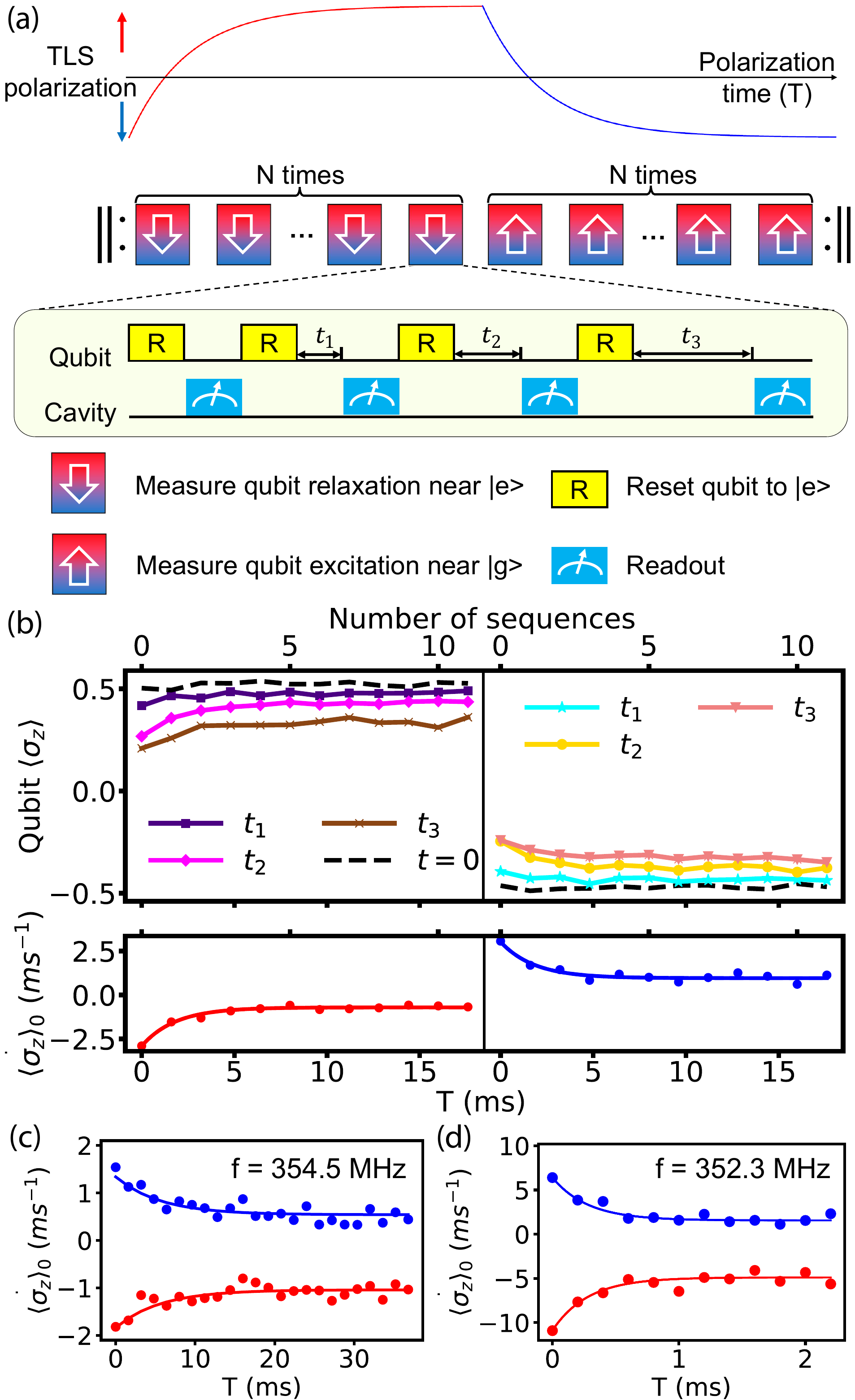}
    \caption{\textbf{Two-timescale relaxometry: protocol and example data.} 
    \textbf{(a)} Schematic of the experiment. In the first half of the cycle we try to repeatedly extract     qubit $\Gamma_\downarrow$ while we polarize TLS towards the excited state over a slow time scale of $T$, and in the second half we do the opposite. Each cycle contains $2N$ sequence blocks, while each block includes 4 qubit resets and readouts. 
    \textbf{(b)} Sample data for extracting the decay slope of qubit polarization, with $N=12$, $t_1=40$ $\mu$s, $t_2=150$ $\mu$s, $t_3=330$ $\mu$s, and block length 1.6 ms.      Top panel shows the raw data of qubit measurements calibrated in units of $\langle\sigma_z\rangle$. Bottom panel shows the extracted initial slope of qubit polarization, $d\langle\sigma_z\rangle/dt$ at $t=0$, as a function of bath-polarization time $T$.  
    \textbf{(c)} Example data showing long environmental response time, measured with increased $N=25$. 
    \textbf{(d)} Example data with strong qubit-TLS coupling.  The block length is reduced to 200 $\mu$s using a permutation structure of delay times $t_1$, $t_2$,... (See Supplementary Section IV for more details.) 
	}\label{fig:exp_protocol}
\end{figure}

\subsection{Two-timescale relaxometry protocol}
To more efficiently measure non-Markovian relaxation dynamics, we present a technique that can probe qubit relaxation together with the environment relaxation on independent time scales. Our measurement protocol is shown in Fig.~\ref{fig:exp_protocol}(a).  Within each experimental cycle, the TLS environment is first polarized towards its excited state over a long timescale ($T$) via interactions with a qubit that is repeatedly re-initialized to $\ket{e}$, and then polarized towards its ground state with the qubit repeatedly re-initialized to $\ket{g}$.  
Throughout this long bath-polarization cycle, we continually monitor instantaneous qubit decay and excitation rates $\Gamma_\downarrow$ and $\Gamma_\uparrow$ by inserting qubit measurements following variable qubit-TLS interaction time $t$. These short time intervals
$t$ act both as the incremental interaction window to allow the qubit to polarize the bath and as a list of variable delay times to probe the instantaneous qubit relaxation rates under the existing bath condition.  

If the state of the TLS bath changes slowly over time, we may group several qubit resets and readouts into a single measurement block (e.g.~with four different $t$, as shown in Fig.~\ref{fig:exp_protocol}(a)), and the bath is treated as in a quasi-steady state within the block.  Within each block, by fitting the qubit state $Z$ as a function of $t$ to an exponential decay curve, we can obtain the initial slope of the qubit relaxation $\dot{Z}_0$, i.e.~the time-derivative of the qubit polarization $dZ/dt$ at $t=0$.  We note that a four-point exponential fit can usually yield $\dot{Z}_0$ robustly even if the timescale of the qubit relaxation dynamics is not known a priori.  This quasi-steady state approximation becomes inaccurate when the qubit dynamics is dominated by a single long-lived TLS.  In such case, we perform only one qubit readout per block but vary its delay time $t$ in an additional layer of sequence loops to allow accurate extraction of $\dot{Z}_0$.  See Supplementary Section III(D) and IV for details of this modified protocol and further discussions of the quasi-steady state approximation of the bath.  

In Fig.~\ref{fig:exp_protocol}(b), we show a typical data set of fitted $\dot{Z}_0$ as a function of the sequence block index, averaged over many bath-polarization cycles.  We also associate each block with its start time within the bath-polarization cycle ($T$).  This $\dot{Z}_0(T)$ data can be heuristically interpreted as that the qubit $\Gamma_\downarrow$ rate starts large and saturates to a smaller value as the TLS is saturated towards the excited state in the first half cycle (red curve), the vice versa for $\Gamma_\uparrow$ rate in the second half cycle (blue curve).  In the limit of perfect qubit reset fidelity, by definition we have $\dot{Z}_0=-2\Gamma_\downarrow$ or $\dot{Z}_0=2\Gamma_\uparrow$ when the qubit is initialized to $\ket{e}$ or $\ket{g}$ respectively. 

When the qubit is tuned to different operating frequencies, we observe TLS bath with different relaxation time scales.  Fig.~\ref{fig:exp_protocol}(c) shows a examples of slow bath dynamics that requires more than 20 ms to saturate.  Fig.~\ref{fig:exp_protocol}(d) shows an example of fast bath dynamics.  It should be noted that this fast bath dynamics is often due to strong relaxation through the qubit and should not be taken as a direct measure of the intrinsic TLS lifetime.

\subsection{Decay rate analysis}
To quantify the qubit relaxation rates under the condition of imperfect qubit and bath polarizations, we independently calibrate the qubit initial polarization $Z_{g}$ and $Z_e$ under our reset protocol, and relate $\dot{Z}_0$ to qubit relaxation rates with
$\dot{Z}_0= -Z_{g/e} \Gamma_{\Sigma} - \Gamma_{\delta}
\label{eq:Z0}$, 
where the subscript $g/e$ depends on which state the qubit is initialized in, $\Gamma_{\Sigma} \equiv \Gamma_{\uparrow} + \Gamma_{\downarrow}$ is commonly known as the inverse of the $T_1$ time, and $\Gamma_{\delta} \equiv \Gamma_{\downarrow} - \Gamma_{\uparrow}$ is the rate difference for the up and down transitions of the qubit.  

Over the long timescale of each bath-polarization cycle, the state of the bath converges between two steady states, a higher-energy state ($H$) closer to excited, and a lower-energy state ($L$) closer to ground.  Without loss of generality, for any observables $O$ of the bath, we can write $O(T) = O_H + (O_L-O_H) f(T)$ for the first half cycle and $O(T) = O_L + (O_H-O_L) f(T)$ for the second half cycle, where $f(T)$ is a decay function from $f(0)=1$ to $f(\infty)=0$.  The qubit relaxation rates, $\Gamma_{\uparrow/\downarrow}$ or $\Gamma_{\Sigma/\delta}$, are proxy observables of the bath, hence we may model:
\begin{equation}
\Gamma_{\delta}(T) =
\left\{
\begin{array}{ll}
\Gamma_{\delta H} + (\Gamma_{\delta L} - \Gamma_{\delta H})e^{-\frac{T}{\tau_e}}, \,\text{first half cycle} \\
\\
\Gamma_{\delta L} + (\Gamma_{\delta H} - \Gamma_{\delta L})e^{-\frac{T}{\tau_e}}, \, \text{second half cycle} 
\end{array}
\right.\label{eq:Gamma_delta}
\end{equation}
where $\Gamma_{\delta H/L}$ represents $\Gamma_\delta$ when the bath is in the H/L state.  Here we have assumed a simple exponential model for the bath dynamics, $f(T)=e^{-T/\tau_e}$, adapted to the general observation and the quality of our data, but this analysis can be generalized to more sophisticated decay models.  Using Eq.~\eqref{eq:Gamma_delta} and its analogous equation for $\Gamma_{\Sigma}(T)$, we can fit $\dot{Z}_0(T)$ to obtain all the bath-dependent qubit relaxation rates $\Gamma_{\Sigma H}$, $\Gamma_{\Sigma L}$, $\Gamma_{\delta H}$, $\Gamma_{\delta L}$, as well as the characteristic time of bath relaxation $\tau_e$ under our measurement protocol.

\begin{figure} [ht]
    \centering
    \includegraphics[width=1\linewidth]{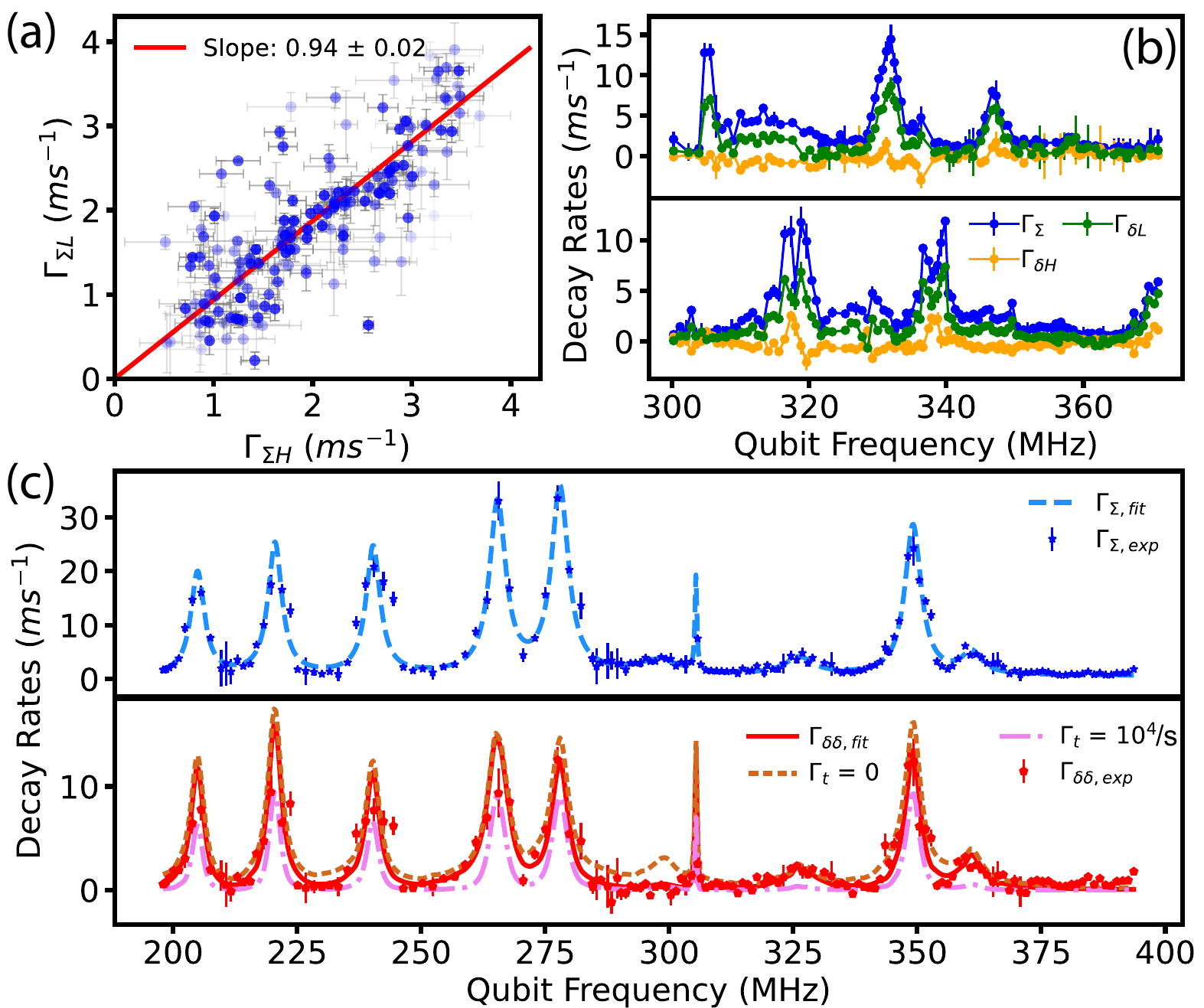}
    \caption{\textbf{Non-Markovian TLS spectroscopy}
    \textbf{(a)} Verification of $\Gamma_{\Sigma}$ being independent of TLS state. Applying the model Eq.~\eqref{eq:Gamma_delta} allows us to obtain a non-Markovian spectroscopy with all four $\Gamma$ rates. Here we plot $\Gamma_{\Sigma L}$ vs.$~\Gamma_{\Sigma H}$ when they are less than 4 ms$^{-1}$ since the uncertainties are significantly higher at large rates. 
    Point transparency is in proportional to its uncertainty. Fitting result of a linear model with intercept fixed to be 0 is given.
    \textbf{(b)} Two example sweeps of non-Markovian relaxation spectroscopy two days apart, which shows significant TLS reconfiguration. 
    \textbf{(c)} Measured relaxation rates from a wider spectroscopy sweep. The upper panel shows a global fit of $\Gamma_{\Sigma}$ with 10 TLS. Each TLS is represented by a Lorentzian. The lower panel includes a global fit to $\Gamma_{\Sigma}$ using our model~\eqref{G_delta_delta_multi-TLS_formula} (solid line) and reference plots of the model-predicted $\Gamma_{\delta\delta}$ under two extreme conditions (dashed lines): All TLS has infinitely long ($\mathrm{\Gamma_{t} = 0/s}$), and 100$\mu$s intrinsic lifetime ($\mathrm{\Gamma_{t} = 10^4/s}$), for comparison. The fact that almost all $\Gamma_{\delta\delta}$ data we showed in the spectroscopy are higher than $\mathrm{\Gamma_{t} = 10^4/s}$ dashed line indicates TLS lifetime of more than 100 $\mu s$. 
	}\label{fig:spec}
\end{figure}

\subsection{Non-Markovian spectroscopy of TLS}
We applied this two-timescale relaxometry protocol to the fluxonium device in a spectroscopy sweep from 200 to 400 MHz. 
We find that the measured results are consistent with $\Gamma_{\Sigma L}\approx\Gamma_{\Sigma H}$ (Fig.~\ref{fig:spec}(a)), indicating an absence of non-Markovian effect from any bosonic bath.  To suppress fit uncertainties from measurement noise, in further analysis we will constrain our model by setting $\Gamma_{\Sigma L}$ = $\Gamma_{\Sigma H}$ = $\Gamma_{\Sigma}$. In Fig.~\ref{fig:spec}(b) we show two sample spectroscopy sweeps that are two days apart, each showing discrete resonance peaks but at different frequencies.  This is consistent with the widely-reported observations that the TLS tend to fluctuate over the timescale of hours to days~\cite{klimov_fluctuations_2018, burnett_decoherence_2019}, and demonstrates our ability to track TLS reconfiguration over a broad spectral range with this technique.

Fig.~\ref{fig:spec}(c) shows another example of processed non-Markovian relaxation spectroscopy, which can be modeled by 10 discrete TLS. 
While $\Gamma_{\Sigma}$ shows how strongly the environment is coupled to the qubit, $\Gamma_{\delta\delta} = \Gamma_{\delta L} - \Gamma_{\delta H}$ reflects the polarizable part of the environment according to our protocol, i.e.~the difference in qubit decay rate between TLS in its high state v.s.~in its low state. Their fraction, $\frac{\Gamma_{\delta\delta}}{\Gamma_{\Sigma}}$, reflects the TLS population difference between its high and low states. The polarizability of TLS is an important sign of its intrinsic lifetime. 

To extract information about these discrete TLS, we apply global fits of the $\Gamma_{\Sigma}$ and $\Gamma_{\delta\delta}$ spectra considering all prominent TLS's contribution. Note that previous qubit $T_1$ spectroscopy studies only contain information on $\Gamma_{\Sigma}$ under the Markovian assumption 
while we aim to study the self- and cross-relaxation rates of the qubit and its TLS environment together. The model is: (see Supplementary Section III for derivations)
\begin{align}
\Gamma_{\Sigma} & = \sum_{k}\Gamma_{qt,k} + \Gamma_q = \sum_{k}\frac{2g_k^2 \Gamma_{2,k}}{\Gamma_{2,k}^2 + \Delta_k^2} + \Gamma_q \label{G_S_multi-TLS_formula} \\
\Gamma_{\delta\delta} &= \sum_{k}\frac{\Gamma_{qt,k}^2 \eta}{\Gamma_{t,k} + \Gamma_{qt,k} \eta} \left[\bar{Z}_{H} - \bar{Z}_{L}\right] \label{G_delta_delta_multi-TLS_formula}
\end{align}
where $\bar{Z}_{H}$ and $\bar{Z}_{L}$ are the time-averaged qubit polarization during the first half (reset to $\ket{e}$) and the second half (reset to $\ket{g}$) of the bath-polarizing cycles,  
$\eta$ is interaction duty-cycle factor of the experiment protocol, defined as the percentage of time within a measurement block that the qubit-TLS interaction is active.  $\eta<1$ because the readout and reset pulse induce significant Stark shift to the qubit that renders its interaction with the relevant (near-resonant) TLS under study inactive.  From Eq.~\eqref{G_S_multi-TLS_formula} we see that $\Gamma_{\Sigma}$ is the sum of a few Lorenzians from discrete TLS and a small background decay rate $\Gamma_q$ that we assume to be a constant for simplicity. $\Gamma_{\delta\delta}$ can be closer to $\Gamma_{\Sigma}$ with longer-live TLS and better qubit initialization, and is closer to 0 if the TLS is short-lived. In Fig.~\ref{fig:spec}(c) we show the results of this multi-TLS fit to $\Gamma_\Sigma$ and $\Gamma_{\delta\delta}$.  For $\Gamma_{\delta\delta}$ we added two reference traces for comparison with the assumption that all TLS have infinite and 100 $\mu$s intrinsic lifetime. All strongly-coupled TLS have their intrinsic lifetime $1/\Gamma_t$ above 100 $\mu$s and many of them in the millisecond range. (We note that a strongly-coupled but short-lived TLS would register a strong peak in $\Gamma_{\Sigma}$) but show no response in $\Gamma_{\delta\delta}$.) The linewidths of the TLS are on the order of few MHz, reflecting their poor phase coherence and/or frequency stability, making the qubit-TLS interaction incoherent. See Supplementary Table II for the extracted TLS parameters. 

We have also briefly carried out two-timescale relaxometry on a different fluxonium qubit in the planar circuit QED architecture.  Here we used quantum non-demolition measurement and FPGA feedback for qubit reset instead of readout-induced reset as in the main device, but all other experimental steps are similar.  This device shows about 12 prominent TLS over the frequency range of 125-325 MHz (See Supplementary Section V(B)).  Most TLS display similar properties as in Fig.~\ref{fig:spec}, i.e.~a high degree of polarizability consistent with millisecond $T_1$ times and linewidths of several MHz.


\subsection{Discussions}
Surface dielectric loss from the TLS bath remains a major limiting factor for coherence times of superconducting qubits today~\cite{wang_surface_2015, siddiqi_engineering_2021}. 
Fluxonium qubits are more protected from dielectric loss than the industry-standard transmon qubits due to their significantly suppressed charge matrix elements, offset by a modestly higher surface participation ratio (SPR) from its smaller shunting capacitor.  Following our standard practice of SPR simulation which treats the TLS bath as a continuum and excludes contributions from the junction oxide and surface layers near the small junctions~\cite{wang_surface_2015}, we find that the fluxonium qubit in this study has SPR about a factor of 4 higher than our typical 3D transmon designs (e.g.~as used in Ref.~\cite{liu_observation_2024}).  This predicts $T_1$ times of about 2 ms assuming a typical surface dielectric quality for our fabrication process with one-step shadow-evaporation as in Ref.~\cite{wang_surface_2015} (a 3 nm thickness surface layer with $\epsilon=10$ and a weighted sum of loss tangent $\tan\delta=2.6\times10^{-3}$), which roughly agrees with the background qubit $T_1$ time $1/\Gamma_q=\mathrm{1.8\pm0.4}$ ms in Fig.~\ref{fig:spec}(b,c). 

However, this excellent background $T_1$ is barely relevant to the average-case performance of the fluxonium as its coherence property is dominated by a forest of discrete TLS resonances with their average spacing only slightly larger than their linewidths.  Previous studies of the fluxonium $T_1$ spectrum~\cite{nguyen_high-coherence_2019, bao_fluxonium_2022}, including Ref.~\cite{somoroff_millisecond_2023} on this very device, provided hints on this dense but individually resolvable TLS population from their large $T_1$ fluctuations versus qubit frequency.  Our use of two-timescale relaxometry finally allows us to disentangle the long lifetimes of these TLS from qubit characterization, resulting in a clean snapshot of the TLS environment. In retrospect, signatures of non-Markovian TLS effect dates back to the pioneering study of 3D fluxonium more than a decade ago~\cite{pop_coherent_2014}, including multi-exponential qubit relaxation and up to 8 ms qubit $T_1$ measured with saturation pulses (which remains unmatched by fluxoniums today and we now believe should be attributed to long TLS relaxation times).

We argue that our observed discrete TLS resonances are not part of the dielectric surface layers in the aforementioned SPR modeling, but rather reside in the oxide tunnel barriers of the junction chain.  Unlike a transmon which consists of only one or two small junctions and statistically tends to encounter no resonant TLS in the junction(s) over a few GHz, the fluxonium consists of nearly 1000 times more volume of junction oxide and can expect no such luck.  Compared to a TLS in the small phase-slip junction which would manifest itself as an avoided crossing in the qubit spectrum (with $g$ on the order of 10 MHz), 
a TLS in an array junction is estimated to have coupling $g$ 100's of times lower, resulting from a combination of the junction chain's voltage division effect and a suppressed charge matrix element (See Supplementary Section VII).  TLS in the surface layers of the capacitor pads and leads of the fluxonium would couple to the qubit too weakly to be individually observable.  The observed density ($\rho=40-60$ $\mathrm{GHz}^{-1}$) and qubit coupling strength (20-80 kHz) of the TLS correspond to an area density of 0.3-0.45 GHz$^{-1}\mu$m$^{-2}$ and effective electric dipole moment of 2-9 Debye for TLS in the junction chain.  Interestingly, these properties are very similar to those reported for TLS in aluminum oxide in the 3-10 GHz range that have been intensively studied in the past~\cite{muller_towards_2019}, suggesting a common origin across nearly two decades of frequency. 

Looking forward, the junction chain quality will likely present a major roadblock to improving qubit coherence through circuit Hamiltonian engineering.  
The measured low background decay rates $\Gamma_q$ and the two-orders-of-magnitude variations of qubit relaxation rate from TLS resonances allows one to reconcile the record-setting fluxonium coherence times in best-case scenarios and the low quality factor of lossy junction oxide averaged over any practical spectral range 
(consistent with previous studies of Al/AlO$_x$ junction chain resonators).  The Josephson junction chain functions as a high-impedance superinductance, which plays a ubiquitous role not only in fluxonium but also in various protected qubits such as the 0-$\pi$ qubits~\cite{gyenis_experimental_2021}, bifluxon qubits~\cite{kalashnikov_bifluxon_2020}, fluxonium molecule~\cite{kou_fluxonium-based_2017}, and a growing family of cos(2$\varphi$) qubits~\cite{smith_magnifying_2022, nathan_self-correcting_2024}.  The relatively low average quality of the junction chain, together with the complex TLS landscape and their resultant non-Markovian dynamics, may have to be considered in the analysis of these novel qubits.  Reducing TLS density in the AlO$_x$ junction chain or finding alternative low-loss superinductors~\cite{gupta_low_2024} appear essential for reliable progress in qubit coherence.  

On the other hand, our study suggests a few interesting directions to investigate and potentially exploit TLS properties.  Our qubit-TLS interaction regime is already in the Purcell regime coveted in the roadmap laid out in Ref.~\cite{odeh_non-markovian_2025}, but strong dephasing of the TLS 
makes it difficult for the coherence of our qubit to benefit from the long $T_1$ times of the TLS.  Since TLS dephasing is believed to be caused by its interaction with slow dynamics of other thermally-activated off-resonant TLS, it will be interesting to explore potential dynamic decoupling~\cite{gustavsson_improving_2013} and spin-locking techniques~\cite{abdurakhimov_identification_2022}.  Furthermore, the fact that our observed TLS density remains comparable to previous literature values despite the much lower frequency and qubit-TLS interaction scale suggests a favorable trade-off to systematically explore even lower frequencies.  Crucially, we argue that our observation of $\Gamma_{\Sigma L}\approx\Gamma_{\Sigma H}$ (also see Ref.~\cite{spiecker_two-level_2023}) suggests that $\Gamma_\Sigma$ is independent of the TLS bath temperature, and hence the Boltzman temperature factor commonly accepted for qubit decoherence analysis~\cite{nguyen_high-coherence_2019, zhang_universal_2021, wang_achieving_2024} should not apply to dielectric loss from TLS, potentially making the heavy-fluxonium regime ($\omega_{01}\ll k_BT$) more favorable than previously appreciated.

Finally, the two-timescale relaxometry demonstrated here provides a general framework for efficient probe of non-Markovian relaxation dynamics of any type of qubits.  Given the ability to reset the qubit on timescale faster than the bath relaxation, it can readily replace traditional $T_1$ measurements in qubit characterization routines, yielding similar measurement throughput while screening and characterizing potential non-Markovoian behavior. \\

\begin{acknowledgments}
\textit{Acknowledgments --} 
We thank E. Dogan for experimental assistance and Y.-Y. Wang for helpful discussions. 
This research is supported by the US Army Research Office, QC-S$^5$ program (No.~W911-NF-23-10093).  Data analysis is partially supported by the US Department of energy, Office of Science, National Quantum Information Science Research Centers, Co-design Center for Quantum Advantage under contract DE-SC0012704.  The planar fluxonium qubit was fabricated and provided by the SQUILL Foundry at MIT Lincoln Laboratory, with funding from the Laboratory for Physical Sciences (LPS) Qubit Collaboratory.
\end{acknowledgments}

\bibliography{Chen_zotero_ref}

\setlength{\textheight}{9.5in}
\setcounter{secnumdepth}{2}

\makeatletter

\definecolor{custompurple}{RGB}{112, 48, 160}
\definecolor{customblue}{RGB}{21, 96, 130}

\makeatother

\setcounter{equation}{0}
\setcounter{figure}{0}
\setcounter{table}{0}
\setcounter{page}{1}
\makeatletter
\renewcommand{\theequation}{S\arabic{equation}}
\renewcommand{\thefigure}{S\arabic{figure}}

\clearpage
\onecolumngrid

\pagebreak
	\begin{center}
	\textbf{\large Supplementary Material for ``Non-Markovian Relaxation Spectroscopy of a Fluxonium Qubit"}
\end{center}

\tableofcontents
\clearpage

\section{Experimental setup}

 The fluxonium device in 3D copper cavity are the same device as presented in Ref.~\cite{somoroff_millisecond_2023}, which also contains detailed fabrication and other related information of the device. Here we list qubit basic information from our measurement (Table \ref{Device_parameters}). Fluxonium's $E_J$, $E_C$ and $E_L$ are extracted by fitting fluxonium's $\ket{g}$-$\ket{e}$ and $\ket{e}$-$\ket{f}$ transitions spectroscopy. Compared to parameters in~\cite{somoroff_millisecond_2023}, $E_C$ is very close while $E_J$ and $E_L$ are both 13\% smaller, which can be well explained by Josephson junctions' aging effect. Cavity $\kappa$ is extracted by fitting the transmission signal to get its full linewidth, while $\chi_{01}$ is obtained by parking the readout tone at the frequency where we observe maximum transmission signal amplitude, and get the angle between $\ket{g}$ and $\ket{e}$ blobs' centers in IQ plane to be 5.4 degrees. Note that as discussed later, we find that normal readout signals are likely to affect qubit states. thus when measuring the angle between $\ket{g}$ and $\ket{e}$, we use very weak readout signal (average photon number less than 1). Though thermal state fluxonium can make the transmission signal full linewidth slightly more broadband than $\kappa$, we think this effect is negligible as $\kappa \gg \chi$. 


\begin{table*}[b]
    \renewcommand{\arraystretch}{1}
    \small
    \begin{tabular}{|c|c|c|c|c|c|c|c|c|c|c|c|c|}
        \hline
        & \multicolumn{3}{|c}{\textbf{Cavity parameters}} & \multicolumn{4}{|c}{\textbf{Fluxonium parameters}} & \multicolumn{5}{|c|}{\textbf{Freq. and coherence at half-flux}} \\ \hline
        & $\omega_{c}/2\pi$ & $\chi/2\pi$ & $\kappa/2\pi$ & $E_J/h$ & $E_C/h$ & $E_L/h$ & Effective qubit& $\omega_{01}/2\pi$ & $\omega_{12}/2\pi$ & $T_1$ & $T_2^*$ & $T_{2E}$\\ 
     & (MHz) & (MHz) & (MHz) & (GHz) & (GHz) & (GHz) & temp. (mK) & (MHz) & (MHz) & ($\mu$s) & ($\mu$s) & ($\mu$s)
        \\ \hline
        3D device & 7519 & 0.41 & 8.7 & 4.88 & 1.09 & 0.56 & 27 & 198 & 4430 & / & 320 & 670 \\ \hline
        Planar device & 6993 & / & 0.2 & 4.08 & 0.92 & 0.35 & $>$100 \footnotemark[1] & 130 & 3825 & / & 78 & 83
        \\
        \hline
    \end{tabular}
    \\
    \footnotemark[1]{The equilibrium qubit population in $\ket{0}$ and $\ket{1}$ appears very close to 50\%:50\% at half flux.} 
    
    \caption{\textbf{Device parameters.} $T_2^*$ and $T_{2E}$ are measured when TLS is not strongly affecting the qubit at half-flux. $T_1$ is not listed as it is better represented by the two-timescale relaxometry data.
    }
    \label{Device_parameters}
\end{table*}

 \begin{figure*}
    \centering
    \includegraphics[width=1\linewidth]{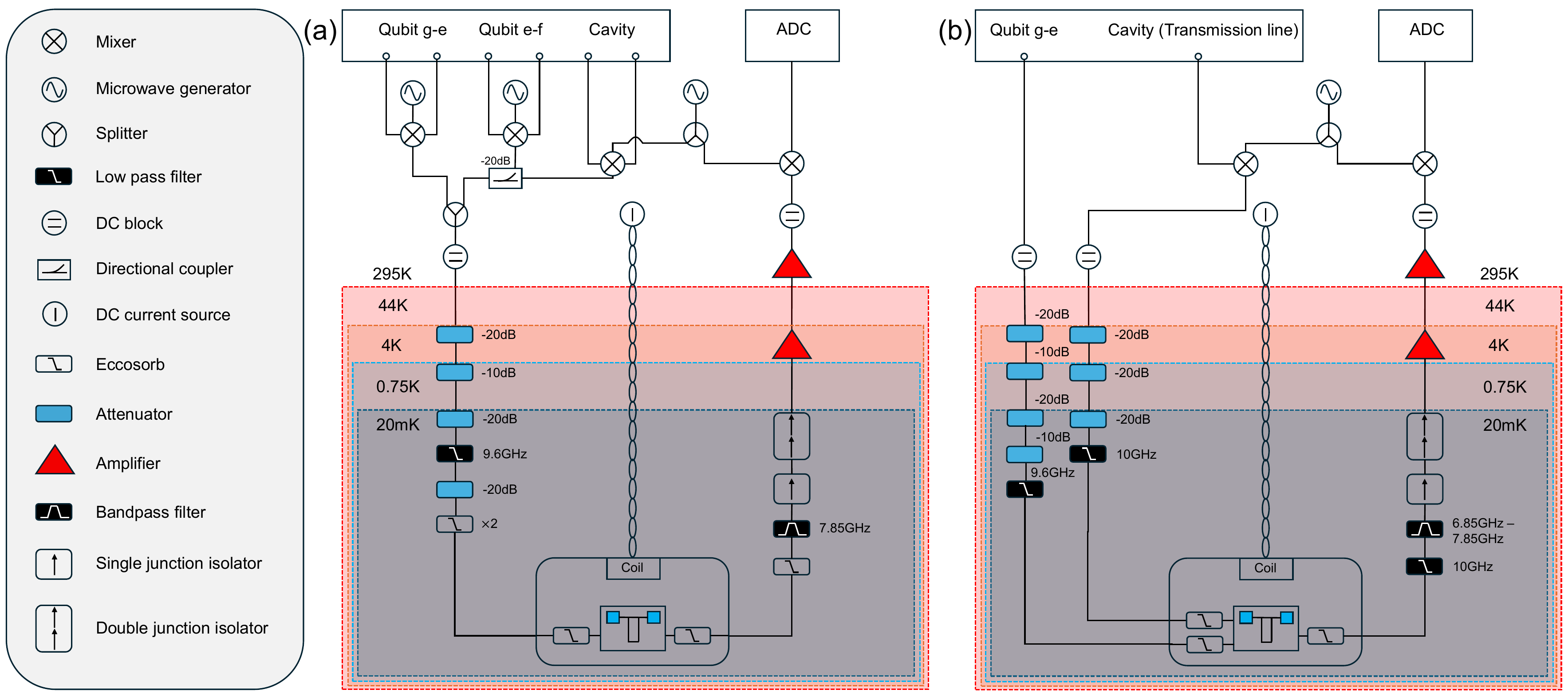}
    \caption{\textbf{Cryogenic and room temperature experimental microwave setup.} \textbf{(a)} 3D device. \textbf{(b)} 2D device.
	}\label{Fridge diagram_combined}
\end{figure*}

The planar device is fabricated and packaged by the SQUILL Foundry at MIT Lincoln Laboratory. It is fabricated on a 350 $\mu$m silicon wafer with a 200 nm base-metal layer of aluminum for large capacitor structures and 30 nm/160 nm double-angle evaporated aluminum for the Josephson junctions. 

Both experiments are carried out in an Oxford Triton 500 Dilution Refrigerator. For the 3D device, We use an amuneal can to protect the device from unwanted external magnetic field. To shield the device from IR radiation, homemade microwave absorber (consists of Stycast 1266, Carbon Black and Silicon Carbide) is painted inside the can and we use eccosorb foam covering inside wall of the can. The whole device is positioned under mixing chamber plate (MXC) that stood a base temperature of around 20 mK. As shown in Fig \ref{Fridge diagram_combined} (a), on the input line, we adopt 70 dB attenuators (40 dB under MXC), a Marki 9.6 GHz low pass filter and three homemade eccosorb (one of them inside the can). For the output line, we apply two homemade eccosorb (one of them inside the can), a Marki 7.85 GHz bandpass filter, a single junction isolator and a double junction isolator to avoid microwave noise entering from output line, followed by a High-Electron-Mobility Transistor (HEMT) amplifier at the 4K stage and a room temperature amplifier. Both RF lines has DC blocks in room temperature setup. We use a DC source (YOKOGAWA 7651) applying current on a superconducting coil to tune flux for the fluxonium qubit. As for the 2D device (Fig. \ref{Fridge diagram_combined} (b)), there are some major changes inside the fridge: Qubit drive line is separated with 60 dB total attenuator attenuation; transmission line also has only 60 dB attenuation; both input line has a low pass filter and one eccosorb; no absorber or eccosorb foam is used.

\section{
Standard ${T_1}$ measurements are susceptible to misinterpretation}
\begin{figure}
    \centering
    \includegraphics[width=0.6\linewidth]{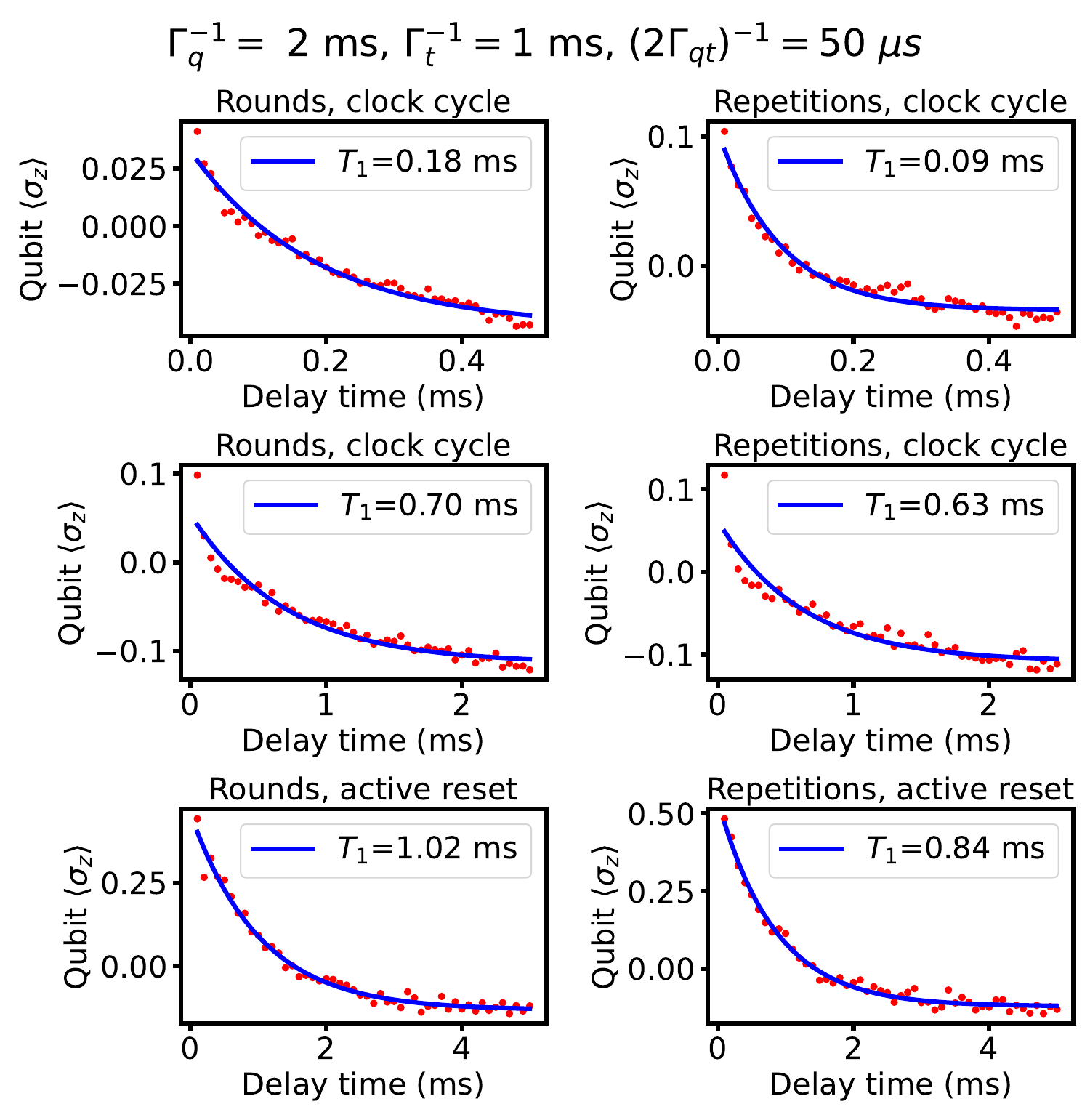}
    \caption{\textbf{Various $T_1$ protocols suffer TLS's impact.} Red dots are data simulated by assigned qubit and TLS parameters together with experiment protocols, and a gaussian-distributed noise is added to the dataset. Blue line is a single exponential fitting to the dataset and so-called ``$T_1$" results are listed. In the left line all measurements are done with rounds and for the right line they are done in repetitions. For the top two figures, we assume the step is 10 $\mu$s and for the middle two figures it is 50 $\mu$s. All four plots are done with clock cycle protocol. As for the bottom two figures, we apply active reset protocol and the step is 100 $\mu$s.}
    \label{regular T1 with TLS}
\end{figure}

In this section, we use some simulated data to illustrate how non-Markovian relaxation dynamics can go unnoticed in a traditional $T_1$ measurement.  The standard exponential fits to decay curves, especially in the presence of noise, and lead to misinterpreted qubit lifetimes that depends on the details of the protocol.  Here we simulate four common scenarios to measure the qubit $T_1$ when a long-lived TLS is unknowingly affecting the qubit relaxation dynamics.  The four scenarios include a combination of two qubit-reset methods (clock cycle and active reset) and two average methods (rounds and repetitions). For clock cycle, we prepare qubit with a qubit $\ket{g}-\ket{e} \pi$ pulse ($\pi_{ge}$) and measure after a variety of delay times, but we keep same cycle period between two neighboring $\pi_{ge}$ pulses.
Active reset, on the other hand, gives very short delay after measurement as we believe there is a robust way to reset qubit to the same state.
Any reset strategy may fall into this category if we don't realize it can only reset the qubit but not the environment. ``Rounds" repeatedly cycles through all measurements by measuring once at each delay time, while ``repetitions" means repeating and averaging measurement results at one delay time before moving to the next.

In the simulation we assume $\mathrm{\Gamma_{qt}=10/ms}$, which leads to a relaxation timescale of 50 $\mathrm{\mu s}$. $Z^{eq}=-0.2$, $p_t^{eq}=-0.1$, and active reset process can always bring qubit Z to 0.5. For each trace we give 50 data points. Readout pulse length is set to be 15 $\mathrm{\mu s}$ and readout process is assumed to be quantum non-demolition, so that each readout result represents average qubit population during readout period. Cycle time for top two figures is 0.5 ms and is 2.5 ms for the middle two ones, while relax time after measurement for active reset protocol is 10 $\mathrm{\mu s}$. $\mathrm{\Gamma_q=0.5/ms}$ and $\mathrm{\Gamma_t=1/ms}$. Reasonable amount of noise is added to all data points. From our simulation results (Fig. \ref{regular T1 with TLS}), $T_1$ measurement by reset method without enough relax delay is very likely to hide non-Markovianity as measured $T_1$ cannot reflect $\mathrm{\Gamma_q}$, $\mathrm{\Gamma_t}$ or $\mathrm{\Gamma_{qt}}$ and it does look like an exponential feature. As for clock cycle protocol, we note that there are also many deceptive results but maybe recognizable if pay attention to short-delay points, and whole contrast value of the curve, which can be much smaller if cycle period or relax delay is not enough. But it is still hard to extract multiple time scales from these protocols without rough estimation of TLS lifetime. We note that fixed delay method, another frequently-used protocol that fixes time between measurement and next qubit preparation $\pi_{ge}$ pulse, shares similar features as clock cycle protocol with our assigned parameters. Overall, our simulation results show that normal $T_1$ measurement protocols can easily cover non-Markovianity and lead to misinterpretation of qubit relaxation dynamics, which needs to be corrected by our proposed method. 

\section{Modeling two-timescale relaxometry} 
\subsection{Decay rate extraction in Markovian environment}\label{Decay rate extraction in Markovian environment}
To introduce concepts in our model, we start from assuming the environment is Markovian. In this case, qubit decay rates $\Gamma_{\downarrow}$ and $\Gamma_{\uparrow}$ are independent of measurement history. Define $Z \equiv \langle \sigma_z \rangle$ for qubit. When qubit is perfectly initialized in $\ket{e}$ state, we note $Z_e^{\mathrm{ideal}} = 1$ and accordingly, $Z_g^{\mathrm{ideal}} = -1$. \\

Qubit decay slope $\frac{dZ}{dt}$ depends on qubit polarization as follow: 
\begin{equation}
\frac{dZ}{dt} = -(1+Z) \Gamma_{\downarrow} + (1-Z) \Gamma_{\uparrow}.\label{general_qubit_decay_rate}
\end{equation}

When reaching equilibrium, the qubit equilibrium polarization $Z^{eq}$ can be expressed by $\Gamma_{\uparrow}$ and $\Gamma_{\downarrow}$:
\begin{equation}
\frac{dZ}{dt} = 0 \Rightarrow Z^{eq} = -\frac{\Gamma_{\downarrow} - \Gamma_{\uparrow}}{\Gamma_{\downarrow} + \Gamma_{\uparrow}}.\label{equilibrium_population}
\end{equation}

Define $\Gamma_{\Sigma}$ and $\Gamma_{\delta}$:
\begin{equation}
\begin{array}{c}
\Gamma_{\Sigma} \equiv \Gamma_{\downarrow} + \Gamma_{\uparrow},\\
\Gamma_{\delta} \equiv \Gamma_{\downarrow} - \Gamma_{\uparrow}.
\end{array}\label{Gamma_sum_and_delta_definition}
\end{equation}


So that: $Z^{eq} = -\frac{\Gamma_{\delta}}{\Gamma_{\Sigma}}$.
Now we can rewrite the qubit decay slope (transition rate) in terms of $\Gamma_{\delta}$ and $\Gamma_{\Sigma}$:

\begin{equation}
\frac{dZ}{dt} = -Z \Gamma_{\Sigma} - \Gamma_{\delta}.\label{qubit_decay_slope_Gamma_sum&Gamma_delta}
\end{equation}

Assume we can initialize our qubit towards two different states, ``excited" and ``ground", with $Z = Z_e$ and $Z = Z_g$, respectively. Qubit state preparation is not ideal, but in general, $Z_e > 0$ and $Z_g < 0$. We can prepare each state, measure the initial slope of the decay, and have two equations to solve two unknowns ($\Gamma_{\Sigma}, \Gamma_{\delta}$):
\begin{equation}
\begin{array}{c}
(\frac{dZ}{dt})_e = -Z_e \Gamma_{\Sigma} - \Gamma_{\delta}\space\space, \\
(\frac{dZ}{dt})_g = -Z_g \Gamma_{\Sigma} - \Gamma_{\delta}\space\space. \label{Slopes_Markovian}
\end{array}
\end{equation}

\subsection{TLS polarizability in Non-Markovian environment}\label{Non-Markovian environment}
Now we assume that the environment depends on a specific measurement sequence. We introduce a new time axis, T, referring to this changing environment, which is different from delay time t in normal $T_1$ measurement. Qubit relaxation and excitation rates can now in some way depend on this variable T. 
$\Gamma_{\downarrow}, \Gamma_{\uparrow} \Rightarrow \Gamma_{\downarrow}(T), \Gamma_{\uparrow}(T)$. All other equations in sub-section \ref{Decay rate extraction in Markovian environment} hold true in this sub-section.\\

We consider numbers of TLS with frequency of $\omega_{tk}$, where k represents the $\mathrm{k^{\text{th}}}$ TLS. We define that $\Gamma_q = \Gamma_{\downarrow, q} + \Gamma_{\uparrow, q}$ and $\Gamma_{tk} = \Gamma_{\downarrow, k} + \Gamma_{\uparrow, k}$ are intrinsic qubit and TLS decay rates, respectively. They are independent from each other and other parameters. $\Gamma_{qtk}$ is qubit-TLS interaction rate, which is a function of qubit frequency $\omega_q$:
\begin{equation}
\Gamma_{qtk}(\omega_q) = \frac{2g^2\Gamma_{2k}}{\Gamma_{2k}^2 + (\omega_{tk} - \omega_q)^2},\label{Gamma_qt}
\end{equation}
where $g$ is qubit-TLS coupling rate and $\Gamma_{2k}$ is qubit and $k^{\text{th}}$ TLS's total dephasing rate (mostly TLS dephasing rate as it is much larger than qubit dephasing rate).\\

Assume $Z$ and $p_{k}$ are qubit and $k^{\text{th}}$ TLS polarizations, respectively, while $Z^{eq}$ and $p_{k}^{eq}$ are thermal equilibrium polarization of qubit and TLS. We then write down the decay slope equations for polarizations of both qubit and TLS as functions of qubit and TLS states~\cite{solomon_relaxation_1955}:


\begin{equation}
\frac{dZ}{dt} = -\Gamma_q (Z - Z^{eq}) - \sum_{k}{\Gamma_{qtk}(Z - p_k)},\label{dpq_dt}
\end{equation}
\begin{equation}
\frac{dp_k}{dt} = -\Gamma_{tk} (p_k - p_k^{eq}) - \Gamma_{qtk}(p_k - Z).\label{dpt_dt}
\end{equation}

Each TLS's quasi-steady-state polarization $p_{k}'$ is found by setting its time derivative $\dot p_k$ to 0:
\begin{equation}
p_{k}' = \frac{\Gamma_{tk} p_{tk}^{eq} + \Gamma_{qtk} p_q}{\Gamma_{tk} + \Gamma_{qtk}}.\label{steady_state_population}
\end{equation}

For our experiment protocol, the polarization sequence consists of constantly initializing the qubit to one of two known states (one is towards $\ket{e}$ and one is close to $\ket{g}$), followed by readout measurements after 4 delay times to extract qubit decay rates. During the measurement sequence, the qubit-TLS interaction is not an invariant: during qubit initialization and readout, as qubit is kicked to out-of-manifold states (not in $\ket{g}$ or $\ket{e}$ states), we believe the interaction is greatly reduced to $\Gamma_{qt}^{\mathrm{eff}}$, the qubit polarization stays at $Z_{\mathrm{pump}}$, and the total percentage of time of these two processes is $\epsilon$. The percentage of the time the interaction is fully on (and equal to $\Gamma_{qt}$) is represented as an efficiency $\eta$. Due to qubit decay during the delay times, the qubit polarization is also not constant during the measurement sequence, and the time-averaged polarization of the qubit during the sequence is the important quantity. When TLS bath is fully polarized, two halves of the measurement sequence (one to polarize the TLS up, and the other to polarize it down) have different average qubit polarizations ($\bar{Z}_{H}$ and $\bar{Z}_{L}$) considering all qubit polarization dynamic details. The lowest and highest TLS polarizations this protocol can reach are calculated out by plugging in time-averaged qubit polarizations during one sequence when bath is fully polarized:

\begin{equation}
p_{k}^{min} = \frac{\Gamma_{tk} p_{tk}^{eq} + \Gamma_{qtk} \bar{Z}_{L}\eta + \Gamma_{qtk}^{\mathrm{eff}} Z_{\mathrm{pump}} \epsilon} {\Gamma_{tk} + \Gamma_{qtk}\eta + \Gamma_{qtk}^{\mathrm{eff}} \epsilon},\label{TLS_min_population}
\end{equation}

\begin{equation}
p_{k}^{max} = \frac{\Gamma_{tk} p_{tk}^{eq} + \Gamma_{qtk} \bar{Z}_{H} \eta + \Gamma_{qtk}^{\mathrm{eff}} Z_{\mathrm{pump}} \epsilon} {\Gamma_{tk} + \Gamma_{qtk}\eta + \Gamma_{qtk}^{\mathrm{eff}} \epsilon}.\label{TLS_max_population}
\end{equation}

This allows us to define each TLS's polarizability: 
\begin{equation}
p_{k}^{\Delta} \equiv p_{k}^{max}  -p_{k}^{min} = \frac{\Gamma_{qtk} \eta}{\Gamma_{tk} + \Gamma_{qtk} \eta + \Gamma_{qtk}^{\mathrm{eff}} \epsilon} \left[\bar{Z}_{H} - \bar{Z}_{L}\right],\label{TLS_population_difference}
\end{equation}
where $\bar{Z}_{H}$ and $\bar{Z}_{L}$ are the time-averaged qubit polarizations during the first half (reset to close to $\ket{e}$) and the second half (reset to close to $\ket{g}$) of the bath-polarizing cycles.

\subsection{Decay rate evolution with changing environment}

So far we have derived how much difference we can make to TLS polarization with our protocol, and TLS state change can lead to qubit decay rate evolution. The environment bath is affected by the bath-polarizing sequences in the following way: when the qubit interacts with a TLS, repeatedly bringing the qubit to its excited/ground state polarizes the TLS to its own high/low state, respectively. Such polarization effect exists even qubit and TLS are far-detuned. Consider the measurement sequence in two halves: In the first half, the qubit is repeatedly brought to its excited state, and its initial transition rate (mostly consists of $\Gamma_{\downarrow}(T)$, but due to imperfection in qubit state preparation, part of $\Gamma_{\uparrow}(T)$ is also included) is measured. Repeated measurements of this type act to polarize the TLS to its excited state (note highest state it can reach as 'H'). As TLS approaches its maximum excited state, $\Gamma_{\downarrow}(T)$ decreases to its minimum. The second half of the measurement is essentially the reverse action: the qubit is repeatedly brought to its ground state, and its transition rate is again measured. The TLS is pushed towards its ground state (note lowest state as 'L'), and $\Gamma_{\uparrow}(T)$ decreases.\\ 

For the first half of the measurement that polarizes TLS from L to H state, the sequence resets the qubit to $\ket{e}$ and both $\Gamma_{\uparrow}$ and $\Gamma_{\downarrow}$ follow an exponential trend, with time constant of $\tau_e$ (note that in principle there should be a sum here for various timescales, but here we assume single environmental timescale, which can describe our experiment data well, especially capture start and end points):
\begin{equation}
\Gamma_{\uparrow} = \Gamma_{\uparrow H} + (\Gamma_{\uparrow L} - \Gamma_{\uparrow H})e^{-T/\tau_e},\label{Gamma_up_env_form}
\end{equation}
\begin{equation}
\Gamma_{\downarrow} = \Gamma_{\downarrow H} + (\Gamma_{\downarrow L} - \Gamma_{\downarrow H})e^{-T/\tau_e}.\label{Gamma_down_env_form}
\end{equation}

Plug these formulas together with \eqref{Gamma_sum_and_delta_definition} into \eqref{Slopes_Markovian}, the initial slope of the qubit decay can then be written as:

\begin{equation}
\frac{dZ}{dt}(\downarrow) = -Z_e \Gamma_{\Sigma H} - \Gamma_{\delta H} - [Z_e(\Gamma_{\Sigma L}-\Gamma_{\Sigma H})+\Gamma_{\delta L} - \Gamma_{\delta H}]e^{-T/\tau_e}.\label{Slope_measuring_gamma_down}
\end{equation}

For the second half, the TLS polarization is reversed, from H to L state. Similarly, the initial slopes are derived to be:

\begin{equation}
\frac{dZ}{dt}(\uparrow) = -Z_g \Gamma_{\Sigma L} - \Gamma_{\delta L} - [Z_e(\Gamma_{\Sigma H}-\Gamma_{\Sigma L})+\Gamma_{\delta H} - \Gamma_{\delta L}]e^{-T/\tau_e}.\label{Slope_measuring_gamma_up}
\end{equation}





From general observations in our experiments, we found that $\Gamma_{\Sigma H} = \Gamma_{\Sigma L}$, so that $\Gamma_{\Sigma}$ is independent of TLS states. Now we can link $\Gamma_{qt}$, $\Gamma_{q}$ to $\Gamma_{\Sigma}$ and $\Gamma_{\delta}$. Recall equations \eqref{general_qubit_decay_rate} and \eqref{dpq_dt}, so that:
\begin{equation}
\Gamma_q(Z-Z^{eq})+\sum_{k}{\Gamma_{qtk}(Z-p_k)}=(1+Z)\Gamma_{\downarrow}-(1-Z)\Gamma_{\uparrow}.
\end{equation}

Separate terms that contain $Z$ and only constant. By comparing the coefficients we can now find that:
\begin{equation}
\Gamma_{\Sigma} = \Gamma_q + \sum_{k}{\Gamma_{qtk}},\label{Gamma_sum_multi-TLS}
\end{equation}

\begin{equation}
\Gamma_{\delta} = -\Gamma_q Z^{eq} -\sum_{k}{\Gamma_{qtk} p_{k}}.\label{Gamma_delta_multi-TLS}
\end{equation}

$\Gamma_{\delta}$ is dependent on TLS polarizations $p_k$. Now consider the upper and lower limit of $\Gamma_{\delta}$, which correspond to environment in its H and L states, respectively. At that time $\Gamma_{\delta}$ is obtained by plugging in all TLS's highest and lowest state polarizations $p_k^{max}$ and $p_k^{min}$:

\begin{equation}
\Gamma_{\delta H} = -\Gamma_q Z-\sum_{k}\Gamma_{qtk} p_k^{max} ,\label{Gamma_delta_H}
\end{equation}

\begin{equation}
\Gamma_{\delta L} = -\Gamma_q Z-\sum_{k}\Gamma_{qtk} p_k^{min} .\label{Gamma_delta_L}
\end{equation}

So that by applying polarization sequences in our experiment, the polarizable part of $\Gamma_{\delta}$ rate is:
\begin{equation}
\Gamma_{\delta\delta} = \Gamma_{\delta L} - \Gamma_{\delta H} = \sum_k{\Gamma_{qtk} p_k^{\Delta}}
=\sum_k{\frac{\Gamma_{qtk}^2 \eta}{\Gamma_{tk} + \Gamma_{qtk} \eta+ \Gamma_{qtk}^{\mathrm{eff}} \epsilon}} \left[\bar{Z}_{H} - \bar{Z}_{L}\right].\label{Gamma_delta_delta_formula}
\end{equation}


\begin{figure}[t]
    \centering
    \includegraphics[width=0.8\linewidth]{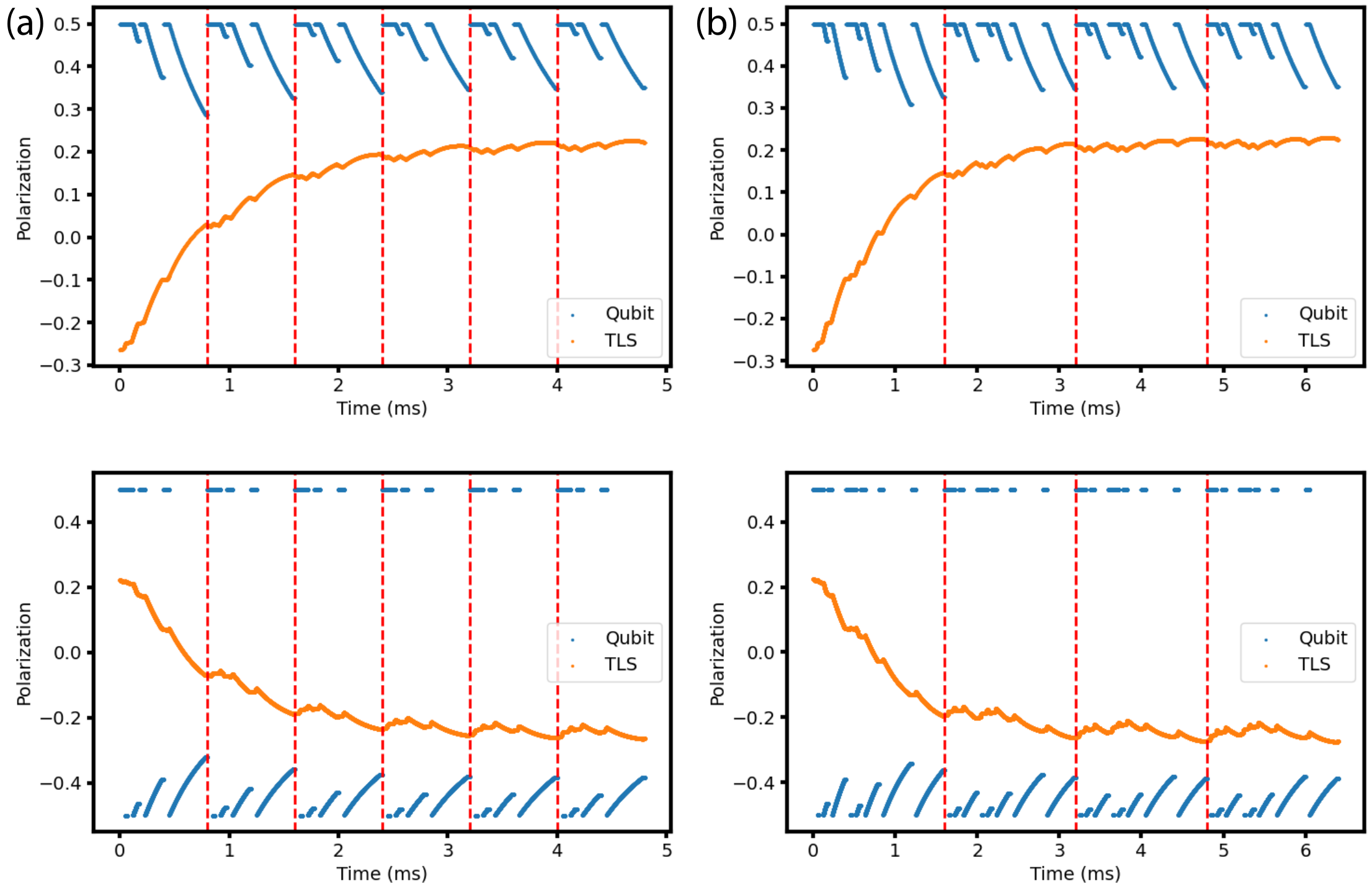}
    \caption{\textbf{Full dynamics analysis of qubit-TLS system.} In the simulation we assume $\Gamma_{qt} = 1$ $\mathrm{ms^{-1}}$, $\Gamma_{q} = 0.5$ $\mathrm{ms^{-1}}$,
    $\Gamma_{t} = 0.5$ $\mathrm{ms^{-1}}$. Qubit is always pumped to close to $\ket{e}$ state with $P_e$ of 0.5. To initialize towards $\ket{g}$, a $\pi_{ge}$ pulse is played at the end of 15 $\mu$s idle time. \textbf{(a)} Here we show qubit and TLS evolution throughout our experiment protocol till states evolution becomes stable. Blue trace represents qubit state change while the orange trace shows TLS state change. Two figures demonstrate up and down polarization process, respectively. Red vertical lines divide time by polarization sequence number, and each sequence is 0.8 ms (detailed composition is in Table \ref{Full dynamics analysis of qubit-TLS system}).  \textbf{(b)} Qubit-TLS full dynamics when we move opposite readout into the closest loop. Now each polarization sequence is 1.6 ms.
    }\label{Full dynamics analysis of qubit-TLS system}
\end{figure}

In the spectroscopy analysis we show in Fig.~\ref{fig:spec}(c), we take $\Gamma_{qt}^{\mathrm{eff}}$ to be 0. We first extract $\Gamma_{qt}$ from $\Gamma_{\Sigma}$ spectrum, and then do analysis to $\Gamma_{\delta\delta}$ rates. $\eta$ is related to our protocol, which will be shown in Section \ref{Two-timescale relaxometry with cycling delay times}. 

\subsection{Choice of protocol parameters and quasi-steady state approximation}
In our non-Markovian relaxation spectroscopy sweep of the device described in the main text, the polarization sequence we play is repeated `initialize - delay - readout' with 4 fixed delay times. We will refer to this two-timescale relaxometry sequence as FD-4.  While the optimal choice of delay times should be tailored to the qubit environment at each flux point, for convenience we used $t_{0,1,2,3}=1$ $\mu$s, 40 $\mu$s, 150 $\mu$s, 330 $\mu$s throughout the whole spectral sweep.  The choice of these delay times are motivated by the goal of sensitively detecting non-Markovian effects when the qubit and environment are both long-lived and coupled weakly to each other (i.e. the background rates away from TLS resonances) while also covering a potentially wide range of potential relaxation rates, from sub 100 $\mu$s to milliseconds. We note that the long environmental response time we often observe is mainly contributed by some less-coupled but polarizable TLS such as those in the junction chain but farther-detuned.

As mentioned in the main text, we apply quasi-steady approximation to TLS state within one measurement block. In this section we will also briefly discuss how qubit and TLS polarizations evolve in the FD-4 experiment. In Fig. \ref{Full dynamics analysis of qubit-TLS system} (a) we demonstrate a full dynamics of qubit-TLS evolution with some common parameters. Admittedly, the TLS state changes quite substantially during the first polarization sequence according to our protocol and it also inevitably fluctuates in the subsequent sequences, but TLS state overall follows an exponential-like pattern.
In this particular run, in order to eliminate potential RO phase drifts, we build in twice the measurements in each sequence block, leading to 1.6 ms rather than 800 $\mu$s blocks (see Section \ref{Calibration experiments}), which makes this approximation a bit worse (Fig. \ref{Full dynamics analysis of qubit-TLS system} (b)). To deal with stronger qubit-TLS interaction cases, we use CD-8 protocol, which will be described in the next section.

\section{Two-timescale relaxometry with cycling delay times} \label{Two-timescale relaxometry with cycling delay times}
\begin{figure*}
    \centering   
    \includegraphics[width=0.7\linewidth]{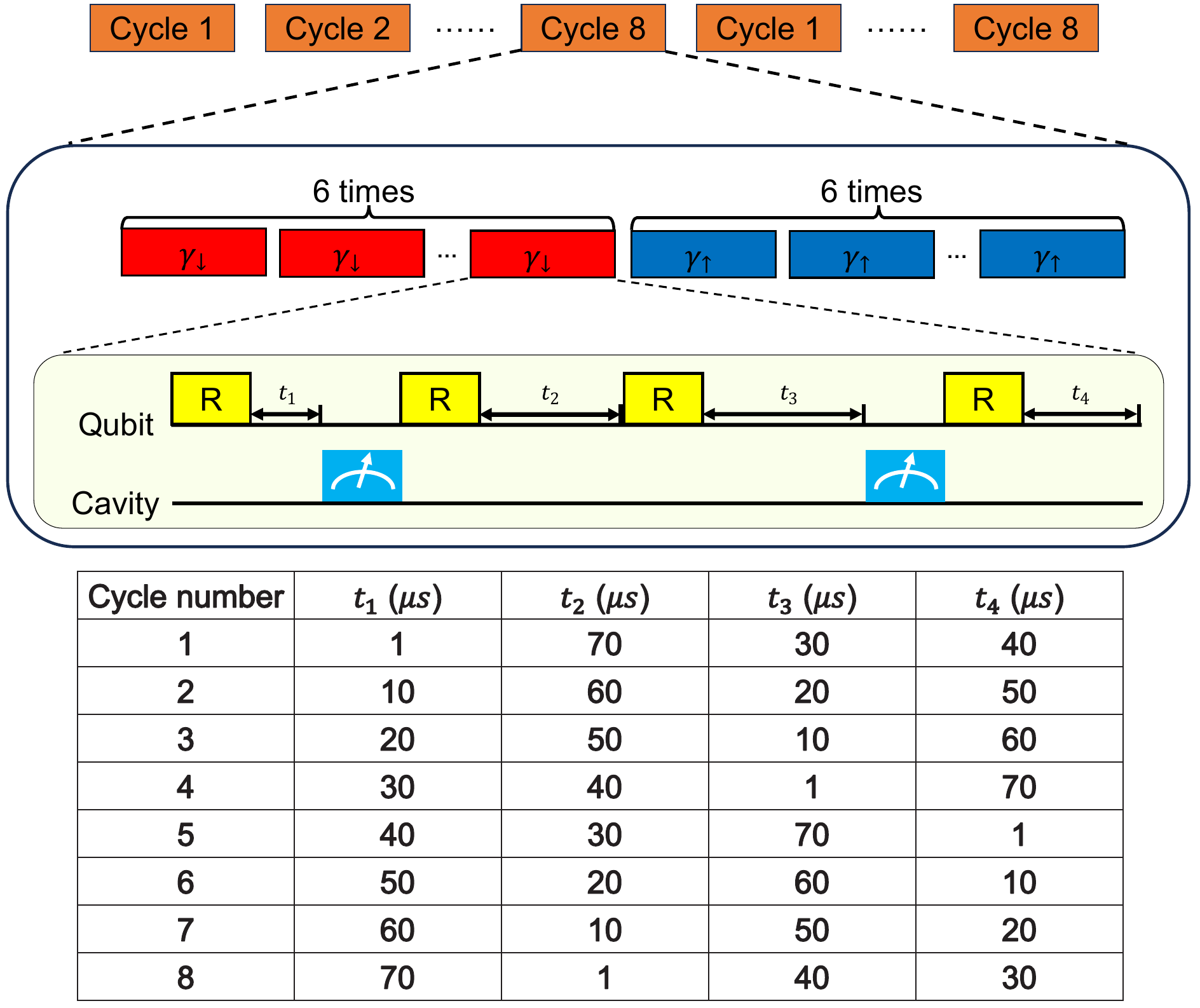}
    \caption{\textbf{Pulse sequence for two-timescale relaxometry with cycling delays (CD-8).} This experiment protocol is similar to the FD-4 experiment (with 4 fixed delays). On top of keep polarizing TLS environment with certain delay times, we loop through different delay time combinations. 
	}\label{Cycle_delay_pulse_sequence}
\end{figure*}

As described in the main text, the main experiment (FD-4) consists of qubit initialization (to either $\ket{e}$ or $\ket{g}$), followed by four different delay times (1 $\mu$s, 40 $\mu$s, 150 $\mu$s and 330 $\mu$s). For those qubit-TLS strong interaction regions, we shorten the delay times to 1 $\mu$s, 10 $\mu$s, 20 $\mu$s, 30 $\mu$s, 40 $\mu$s, 50 $\mu$s, 60 $\mu$s and 70 $\mu$s. Note that 1 $\mu$s delay time in principle should be 0 but set to be 1 $\mu$s due to technique issue. As this experiment is more time-consuming, in order to combine it with our regular protocol, we first do a FD-4 experiment and extract its $\Gamma_{\Sigma}$ result. If $\Gamma_{\Sigma}$ is larger than a threshold number (4 $\mathrm{ms^{-1}}$ for the spectroscopy shown in the main text), we do a CD-8 experiment and the result can overwrite FD-4 experiment's in spectroscopy for this flux point. Detailed pulse sequence is shown in \ref{Cycle_delay_pulse_sequence}. We keep looping through cycle 1 to 8 and in each cycle, we polarize TLS to high state by six sequences and do the opposite also for six times. Four delay times $t_1$, $t_2$, $t_3$ and $t_4$ vary with cycle according to the table. Readouts are only set after first and third delay times in each sequence and delay time order is carefully chosen so that between every two readouts, environment have same effective interaction time with the qubit. Data analysis is carried out in two steps, similar to the FD-4 experiment, but the first step is an eight-point exponential fitting using all readouts in $t_1$ column and $t_3$ column in the same sequence index. Thus in the second step, there are twelve $\frac{d Z}{dt}$ data points for each polarization direction. As mentioned in the previous section, each polarization sequence in the FD-4 experiment stands for 1.6 ms, and here the neighboring slope measurement interval is 200 $\mu$s.

In the FD-4 experiment, we assume qubit initialization states remain unchanged for same pumping scheme and qubit frequency, meaning it is independent of TLS state.  This is a good approximation when the qubit-TLS interaction is not very strong.  For the CD-8 experiment intended for the regime of strong qubit-TLS interaction, we do see qubit initializations vary slightly with TLS state. This is because the TLS affects the qubit state during the short idling time in our reset protocol. Thus, $Z_e$ and $Z_g$ are functions of T. Our slope functions become:
\begin{equation}
\begin{array}{c}
(\frac{dZ}{dt})_e = -Z_e(T) \Gamma_{\Sigma} - \Gamma_{\delta}\space\space, \\
(\frac{dZ}{dt})_g = -Z_g(T) \Gamma_{\Sigma} - \Gamma_{\delta}\space\space. \label{Slopes_Markovian}
\end{array}
\end{equation}

Assume that $Z_e(T)$ and $Z_g(T)$ both evolve exponentially with T, and time constant is the same as $\tau_e$. General observation indicates that contrast of two initializations are equal, so $\delta P_e$ = $\delta P_g$ = $\delta P$. Then slope functions become:
\begin{equation}
\begin{array}{c}
\frac{d Z}{dt}(\downarrow) = -P_{\ket{e}} \Gamma_{\Sigma} - \Gamma_{\delta H} - [(\Gamma_{\delta L} - \Gamma_{\delta H})e^{-T/\tau_e}] - \delta P \Gamma_{\Sigma} e^{-T/\tau_e}, \\
\\
\frac{d Z}{dt}(\uparrow) = -P_{\ket{g}} \Gamma_{\Sigma} - \Gamma_{\delta L} - [(\Gamma_{\delta H} - \Gamma_{\delta L})e^{-T/\tau_e}] - \delta P \Gamma_{\Sigma} e^{-T/\tau_e}. \label{slope_equation_both_cycles}
\end{array}
\end{equation}

\begin{table} [h]
    \centering
    \renewcommand{\arraystretch}{1.5}
    \begin{tabular}{|c|c|c|c|c|c|c|c|c|c|}
        \hline
        Protocol & Pumping & \textcolor{red}{Idle time} & \textcolor{red}{$t_1$} & \textcolor{red}{$t_2$} & \textcolor{red}{$t_3$} & \textcolor{red}{$t_4$} & Readouts & Extra waiting period & Total
        \\ \hline
        FD-4 & 35 $\mu$s $\times$ 4 & \textcolor{red}{15 $\mu$s $\times$ 4} & \textcolor{red}{1 $\mu$s} & \textcolor{red}{40 $\mu$s} & \textcolor{red}{150 $\mu$s} & \textcolor{red}{330 $\mu$s} & 15.4 $\mu$s $\times$ 4 & 17.4 $\mu$s & 800 $\mu$s
        \\ \hline
        CD-8 & 35 $\mu$s $\times$ 2 & \textcolor{red}{15 $\mu$s $\times$ 2} & \multicolumn{4}{c|}{\textcolor{red}{Total delay time: 70 $\mu$s}} & 15.4 $\mu$s & 14.6 $\mu$s & 200 $\mu$s
        \\ \hline

    \end{tabular}
    \caption{\textbf{Time spent on each stage of two protocols.} Red parts are counted into effective interaction portion.}
    \label{Time spent on each stage of two protocols.}

\end{table}

$\Gamma_{\Sigma}$ and $\Gamma_{\delta\delta}$ formulas are the same as \eqref{Gamma_sum_multi-TLS} and \eqref{Gamma_delta_delta_formula}, respectively. So the following analysis is the same as our FD-4 experiment. As described before, according to our experiment protocol, the duty cycle efficiency $\eta$ is 73\% for FD-4 and 50\% for CD-8 (details in Table \ref{Time spent on each stage of two protocols.}).

\section{Additional data}
\subsection{Relaxation spectroscopy of the 3D fluxonium device (data associated with Fig.~\ref{fig:spec}(c))}

\begin{table*}[b]
    \centering
    \renewcommand{\arraystretch}{1.5}
    \resizebox{\textwidth}{!}{%
    \begin{tabular}{|c|c|c|c|c|c|c|c|c|c|c|}
        \hline
        TLS number & 1 & 2 & 3 & 4 & 5 & 6 & 7 & 8 & 9 & 10 \\
         \hline
        g/(2$\pi$) (kHz) & 50 $\pm$ 2 & 57 $\pm$ 3 & 57 $\pm$ 3 & 70 $\pm$ 5 & 75 $\pm$ 3 & 32 $\pm$ 5 & 18 $\pm$ 12 & 30 $\pm$ 4 & 66 $\pm$ 2 & 28 $\pm$ 2
        \\ \hline
        $\Gamma_2/(2\pi)$ (MHz) & 1.7 $\pm$ 0.2 & 1.7 $\pm$ 0.3 & 1.7 $\pm$ 0.2 & 2.0 $\pm$ 0.4 & 2.1 $\pm$ 0.3 & 5.3 $\pm$ 2.0 & 0.2 $\pm$ 0.2 & 3.4 $\pm$ 1.3 & 2.0 $\pm$ 0.2 & 2.4 $\pm$ 0.5 \\ \hline
        Frequency (MHz) & 204.9 $\pm$ 0.2 & 220.5 $\pm$ 0.2 & 240.3 $\pm$ 0.3 & 265.4 $\pm$ 0.3 & 278.0 $\pm$ 0.2 & 297.8 $\pm$ 1.1 & 305.4 $\pm$ 0.2 & 326.2 $\pm$ 1.0 & 349.1 $\pm$ 0.1 & 361.0 $\pm$ 0.3 \\ \hline
        $\Gamma_t$ ($\mathrm{ms^{-1}}$) & 0.6 - 1.5 & 0.8 - 2.0 & 0.1 - 1.1 & $<$ 0.1 & 1.8 - 4.2 & $\sim$10 & 1.1 - 3.6 & $<$ 0.1 & 1.2 - 2.2 & $<$ 0.1 \\ \hline
        
    \end{tabular}
    \caption{\textbf{TLS information for Fig. \ref{fig:spec}.} Model we used for extracting these parameters are described in section \ref{Non-Markovian environment}. We note that for TLS \#7 the fitting result has very large uncertainties as the peak is mostly composed of a single point, but this peak shows up in back and forth sweeps both times and hence is confirmed as a discrete TLS.}
    \label{TLS_fitting_result}
    }
\end{table*}

\begin{figure}[t]
    \centering
    \includegraphics[width=1\linewidth]{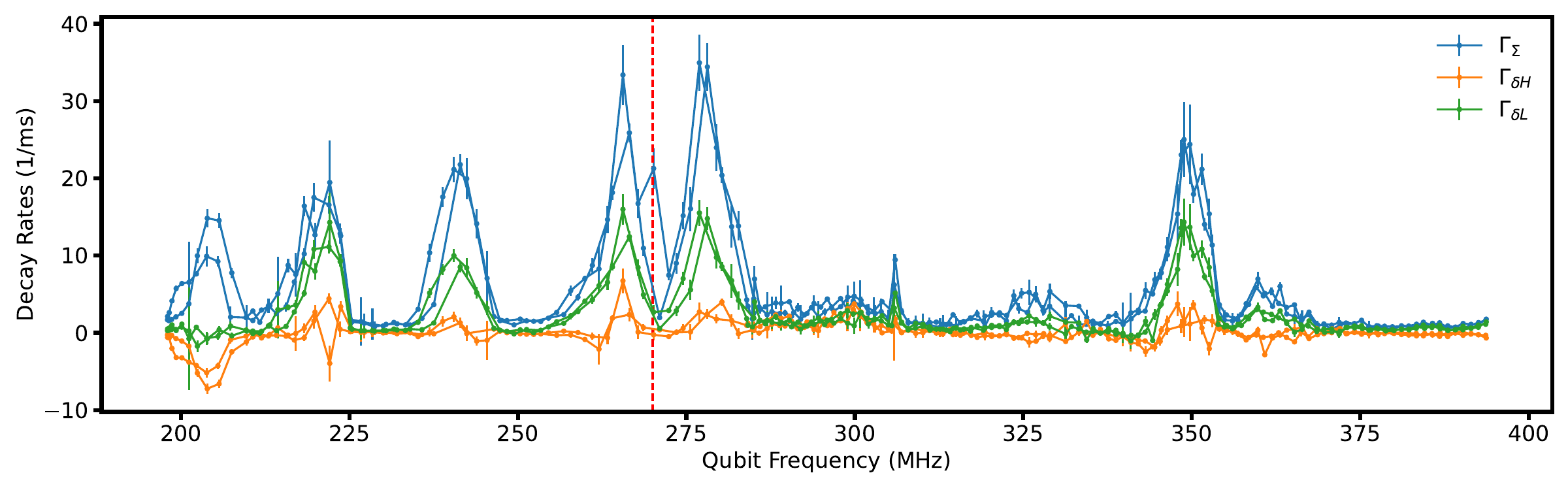}
    \caption{\textbf{Relaxation spectroscopy of the 3D fluxonium device.} Source data of main text Fig. 4(c), which includes a roundtrip sweep starting from the low end of the frequency range. Red dashed line separates the qubit frequency into two parts. For the higher part, two runs have good agreement, thus we take average of two runs. As for lower part, we choose data from the second run. 
}\label{Spectroscopy_original_main_text}
\end{figure}

\begin{figure}
    \centering
    \includegraphics[width=0.6\linewidth]{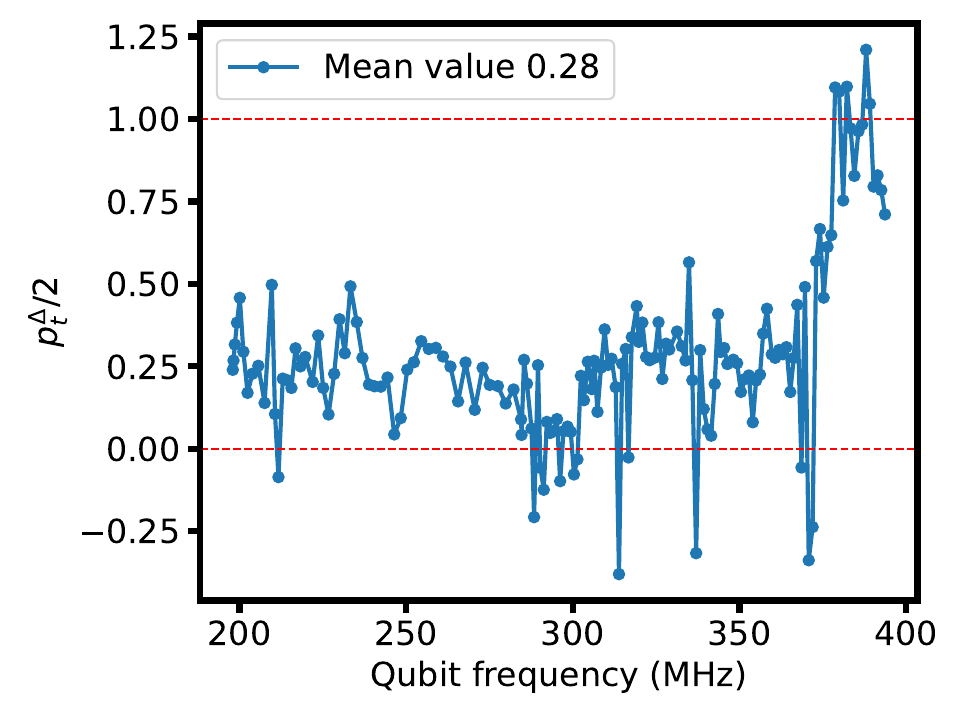}
    \caption{\textbf{TLS polarizability spectrum.} The polarizability is calculated by $\frac{\Gamma_{\delta\delta}}{2(\Gamma_{\Sigma}-\Gamma_q)}$, so that it should range from 0 to 1. Data is averaged for two sweeps and averaged polarizability is 0.28. 
	}\label{Delta_P_t}
\end{figure}

Here we show some relevant results using the analysis above for the spectroscopy shown in main text: In Table~\ref{TLS_fitting_result} we present detailed results of all TLS parameters, and Fig. \ref{Spectroscopy_original_main_text} contains decay rates of both sweeps, which take around 36 hours to complete. Frequency sweep started from half-flux, went up to 400 MHz and came back down. Two sweeps have great consistency above 270 MHz and some discrepancy below that due to TLS reconfiguration. TLS polarizability is displayed in Fig.~\ref{Delta_P_t}, which is calculated by $\frac{\Gamma_{\delta\delta}}{2(\Gamma_{\Sigma}-\Gamma_q)}$ under the assumption that qubit is only affected by a dominant TLS at any frequency point.

\subsection{Relaxation spectroscopy of the planar fluxonium device}
\begin{figure}
    \centering
    \includegraphics[width=0.8\linewidth]{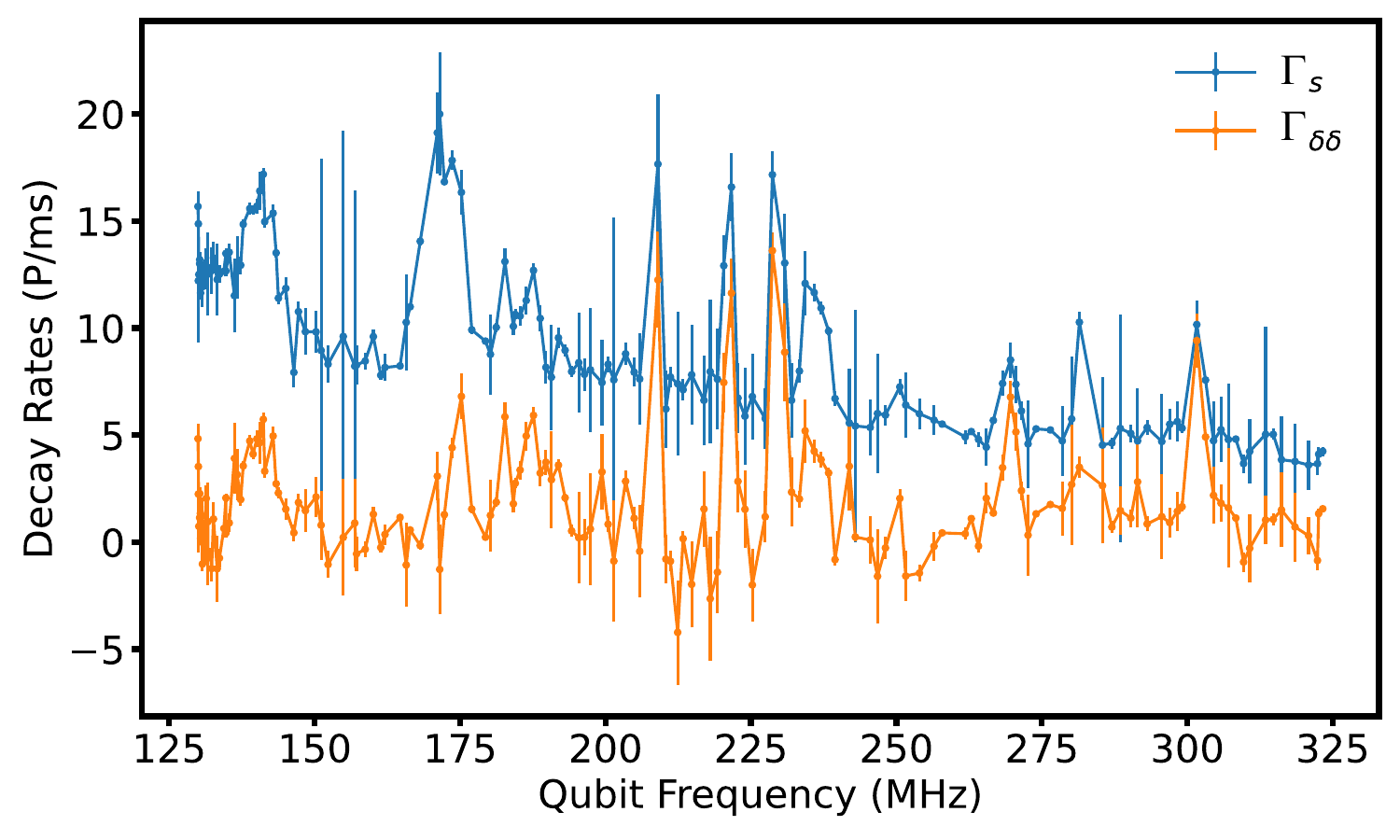}
    \caption{\textbf{Sample spectroscopy of a fluxonium planar device.}}
    \label{Tanvir_TLS_exp_spec}
\end{figure}

The long TLS lifetimes observed in the main device in this study and in Refs.~\cite{spiecker_two-level_2023, liu_observation_2024}) are somewhat surprising given that these devices do not have any engineered phononic structures to shield them from phonon-mediated decays (in contrast to Ref.~\cite{chen_phonon_2024, odeh_non-markovian_2025}).  Coincidentally, all these experiments are done in a 3D circuit QED device architecture, where the qubit chip is fully enclosed by a 3D cavity and only makes edge contact to the cavity package.  To investigate whether the long TLS lifetime may be related to any accidental phononic environment of a 3D cavity packaging, we further applied the non-Markovian relaxation spectroscopy method to a fluxonium in a planar device. 

This device contains a junction chain with 300 large Josephson junctions, each 200 nm in width and 3 $\mathrm{\mu m}$ in length. We employ our FD-4 protocol with four delay times 0 $\mathrm{\mu s}$, 15 $\mathrm{\mu s}$, 30 $\mathrm{\mu s}$ and 50 $\mathrm{\mu s}$. Qubit initialization is realized by measurement-feedback-based active reset with a reset fidelity of over 90\% at half-flux. We observe approximately 12 discrete observable TLS in this spectroscopy (Fig.~\ref{Tanvir_TLS_exp_spec}), corresponding to a TLS density of 0.33 $\mathrm{/\mu m^2/GHz}$. Note that due to large number of junctions in the junction chain and high base decay rate ($\Gamma_q$), some features of discrete TLS in the junction chain may be difficult to solve from the background, so the actual TLS density tends to be larger.  Analysis applied to this spectroscopy indicates that most TLS intrinsic lifetime vary from a few hundred microseconds to a few milliseconds.
Therefore, this results shows that low-frequency TLS generally have long lifetimes in the phononic environment of regular planar qubits as well.

The spectroscopy reveals a Markovian background of qubit $T_1$ which is about 250 $\mu$s at around 300 MHz and decreases to about 80 $\mu$s near half flux (130 MHz).  This background $T_1$ is substantially lower than the 3D device, although the surface participation ratio in this device is only modestly higher.  We attribute this background $T_1$ primarily to inductive loss due to non-equilibrium quasiparticles, which is consistent with the frequency dependence of this background $T_1$.

\section{Calibration experiments} \label{Calibration experiments}
\begin{figure}[b]
    \centering
    \includegraphics[width=0.55\linewidth]{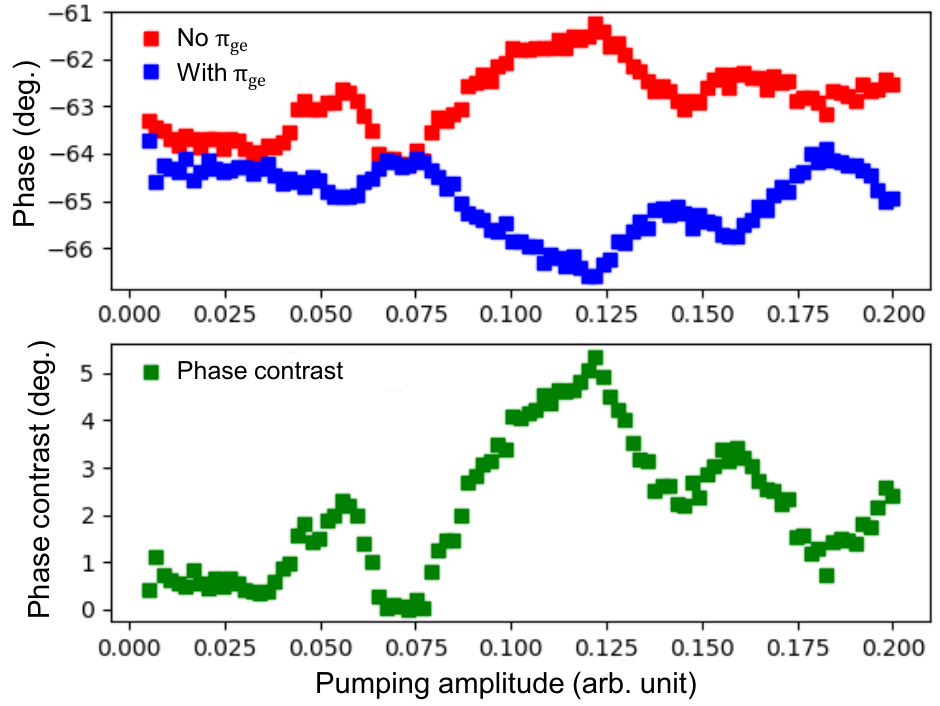}
    \caption{\textbf{Example data of tuning up qubit reset via measurement-induced state transition.} 
    Measured qubit state after a 35 $\mu$s pumping pulse at the readout frequency, as a function of the amplitude of the pumping pulse.  Top panel shows the qubit state (in unit of transmission phase angle) measured directly after the pump pulse, and after an additional $\pi$-pulse.  The difference of the two traces are plotted in the bottom panel, which is proportional to $\langle \sigma_z \rangle$ of the qubit.  
    Here, applying a reset pulse with an optimal amplitude of 0.122 can provide around 5 times qubit polarization than thermal equilibrium.
    }\label{cooling_sweep_sample_data}
\end{figure}

\begin{figure}
    \centering
    \includegraphics[width=0.6\linewidth]{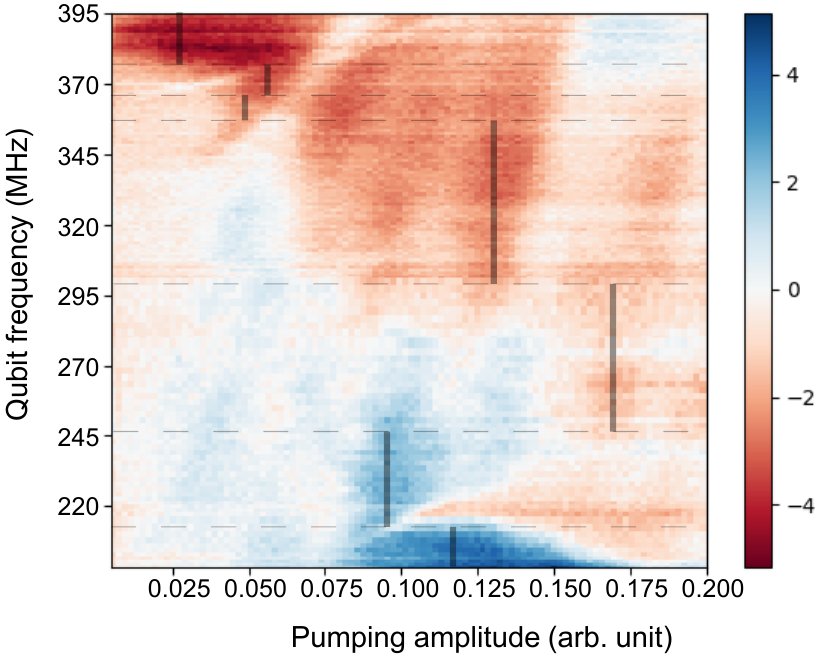}
    \caption{\textbf{Spectroscopy sweep of measurement-induced state transition for determining reset parameters.} After a pumping pulse with various amplitude at the readout frequency, phase contrast between direct readout and readout after a $\pi_{ge}$ pulse is measured, as shown in the color bar, which is proportional to the qubit polarization. Light black solid line roughly shows our chosen pumping amplitude for different frequency period.
	}\label{fig:pumping}
\end{figure}

Due to phase drift happen in our measurement from time to time, all our measurements are done in pairs: one with a qubit $\pi_{ge}$ pulse right before readout, and one without it. The difference in the measured phase of these two readouts is recorded. To have better ability to polarize TLS and reset qubit quickly and repeatedly, we need to prepare qubit first so that they can have larger contrast between their initialization to near $\ket{e}$ and near $\ket{g}$ states, and the reset process should have minimal impact on TLS. Our qubit initialization process utilizes an effect commonly known as readout-induced state transitions, a strong readout may unconditionally pump the qubit to a specific state~\cite{sank_measurement-induced_2016, dumas_measurement-induced_2024, bista_readout-induced_2025}.  
We apply a pumping tone at cavity frequency, followed by some idling time, to reset the qubit to near $\ket{e}$ or near $\ket{g}$ states.  This process can be understood as cavity is excited to some high energy states, and swapped excitations with fluxonium's high energy states. The later decay back to $\ket{e}$ or $\ket{g}$ state with preference in a short time. As shown in Fig.~\ref{cooling_sweep_sample_data}, after doing pumping with certain strength, phase contrast of with and without a $\pi_{ge}$ pulse before the readout alters, indicating that varying pumping strength can affect qubit's final state in $\ket{g}-\ket{e}$ manifold. Choosing pumping amplitude that results in larger phase contrast can initialize our qubit well. Fig.~\ref{fig:pumping} is a sample spectroscopy showing pumping power versus phase contrast and how we choose pumping amplitude for different qubit frequencies.  The very complex landscape of measurement-induced state transition implied in Fig.~\ref{fig:pumping} is not well understood, but a recent study~\cite{bista_readout-induced_2025} suggests that it may in fact be also related to the TLS in the device. Note that in this spectroscopy for above 380 MHz, phase contrast is high even with very low pumping amplitude, which is because previous readout plays a role in resetting the qubit.

During a readout, the ground and first excited states of the fluxonium correspond to the transmitted resonator phases $\phi_g$ and $\phi_e$, respectively (Note that our readout is probably strong enough to excite qubit to higher energy state, but ground and first excited states still have their unique transmitted resonator phases). Then with the assumption that $p_g + p_e = 1$ (all higher states of the fluxonium are unpopulated), we can write direct readout result $M_0$ as $M_0 = p_g \phi_g + p_e \phi_e$, and with a $\pi_{ge}$ before readout, the result becomes $M_{\pi}$, where $M_{\pi}=p_e \phi_g + p_g \phi_e$. Now we find that:
\begin{equation}
M_0 - M_{\pi} = (p_e - p_g) (\phi_e - \phi_g).\label{M_0-M_pi}
\end{equation}

So, as long as we know the transmitted resonator phase contrast ($\phi_e$-$\phi_g$), we can obtain $\ket{g}$ and $\ket{e}$ state populations by measuring the phase contrast of two readouts ($M_0 - M_{\pi}$).\\

As we cannot guarantee perfect qubit state preparation, we design an experiment with four measurement sequences to obtain the phase contrast $\phi_e - \phi_g$. Assume that after pumping at resonator frequency, qubit consists of three states: $\ket{g}$, $\ket{e}$ and $\ket{f}$. $\ket{f}$ represents out-of-manifold state. $\phi_f$ and $p_f$ are the resonator phase and the population of this out-of-manifold state, respectively. All following measurements are done with different delay times after pumping. Here we list four measurement sequences: \\
(1) Pumping - delay - Readout,\\
(2) Pumping - delay - $\pi_{ef}$ - Readout,\\
(3) Pumping - delay - $\pi_{ge}$ - Readout,\\
(4) Pumping - delay - $\pi_{ge}$ - $\pi_{ef}$ - Readout.

Four measurement results $M_1$, $M_2$, $M_3$ and $M_4$ can be expressed by all states' resonator phase and populations, assuming all $\pi$ pulses are perfect:
\begin{equation}
\begin{pmatrix}
    M_1\\
    M_2\\
    M_3\\
    M_4
\end{pmatrix} =
\begin{pmatrix}
p_g & p_e & p_f \\
p_g & p_f & p_e \\
p_e & p_g & p_f \\
p_e & p_f & p_g
\end{pmatrix}
\begin{pmatrix}
\phi_g \\
\phi_e \\
\phi_f 
\end{pmatrix}.\label{Four equations}
\end{equation}

\begin{figure}
    \centering
    \includegraphics[width=0.6\linewidth]{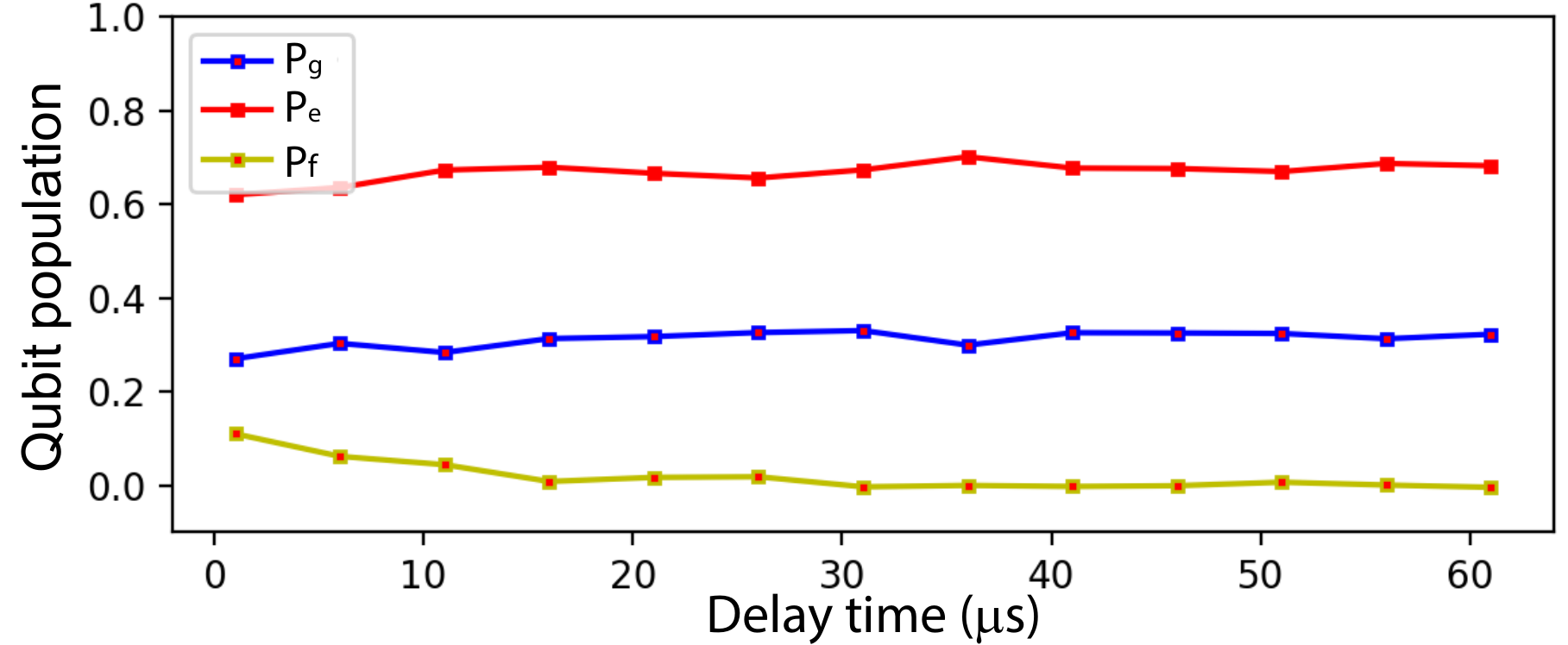}
    \caption{\textbf{Example data of calibrated qubit population distribution following a reset pulse.} At each flux point, to calibrate our qubit reset fidelity and the $\ket{g}$-$\ket{e}$ phase contrast of our readout, we perform four different measurements following qubit reset.  After we compute $\phi_g$, $\phi_e$ and $\phi_f$ as discussed in text (Eq.~\eqref{Four equations}), we can convert measured phase signals at different delay time into population distributions. The data shows that the reset pulse may excited the qubit to higher excited states (especially when the chosen pumping amplitudes is high), hence we incorporate a 15 $\mu$s wait time in our reset sequence to ensure that the out-of-manifold population ($p_f$) is mostly eliminated.
	}\label{Phase_contrast_sample_data}
\end{figure}

\begin{figure}
    \centering
    \includegraphics[width=0.5\linewidth]{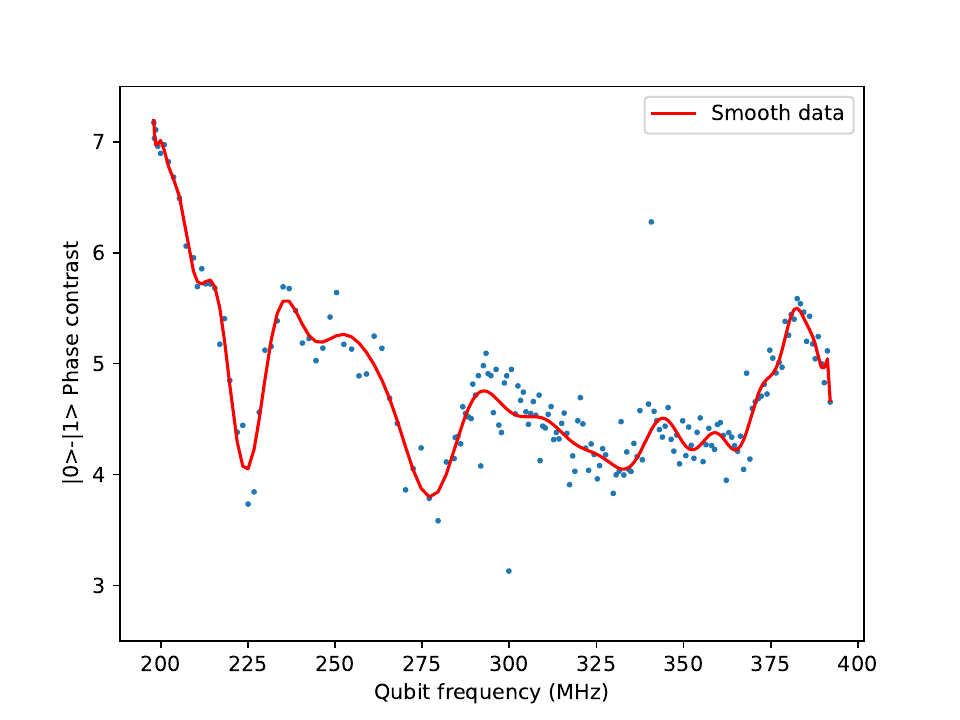}
    \caption{\textbf{Sample data for phase-population coefficient table.} To convert phase contrast of our readouts to qubit population, we need to know what is $\ket{g}$ state to $\ket{e}$ state full phase contrast. We use a polynomial function to smooth this dataset. 
	}\label{Phase_contrast_vs_frequency}
\end{figure}

\begin{figure}
    \centering
    \includegraphics[width=0.6\linewidth]{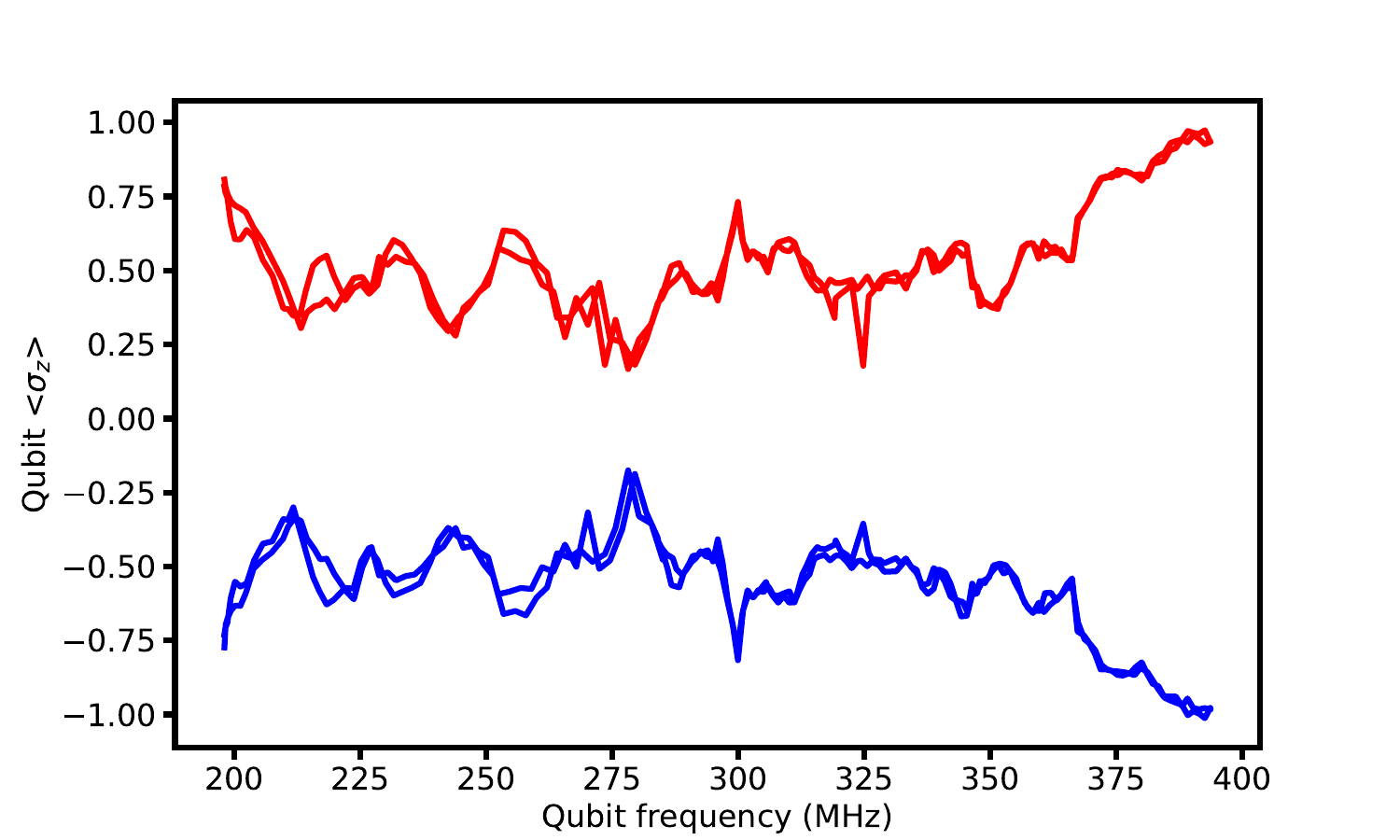}
    \caption{\textbf{Qubit initialization performance in practice.}  Red trace corresponds to $\ket{e}$ state initialization and blue trace shows $\ket{g}$ state initialization. Back and forth sweeps have good agreement.
	}\label{Qubit_initialization_in_exp}
\end{figure}

As shown in Fig. \ref{Phase_contrast_sample_data}, we notice that out-of-manifold state decay fast and after a short period of time, all four readouts become stable, indicating elimination of out-of-manifold state. So we can assume that for our measurements with longest delay time, qubit does not stay in $\ket{f}$, i.e. $p_f = 0$, $p_g + p_e = 1$. With all these equations and assumptions, we can solve all unknown parameters and thus know the resonator phase contrast when qubit is at $\ket{1}$ and $\ket{0}$, $\phi_e-\phi_g$. Data shows that generally if we apply 15 $\mu$s idle time after pumping, out-of-manifold qubit state can be mostly back to $\ket{0}$ and $\ket{1}$ manifold. Thus in our experiment, qubit initialization process includes 35 $\mu$s pumping pulse and 15 $\mu$s idle time. Note that sample data shown in Fig. \ref{cooling_sweep_sample_data} and \ref{fig:pumping} also applies this initialization process. Fig. \ref{Phase_contrast_vs_frequency} provides the calculated ($\phi_e-\phi_g$) phase contrast versus qubit frequency for the spectroscopy sweep in Fig. \ref{fig:spec} (c). Fig. \ref{Qubit_initialization_in_exp} demonstrates how our pumping strategy performs in the experiments that produce spectroscopy shown in main text. Each data point is the average of all readouts for corresponding initialization in two-timescale relaxometry experiment.

\section{Surface participation ratio (SPR) simulation and TLS electric dipole moment}
\subsection{SPR simulation} \label{SPR simulation}
In this sub-section we discuss our SPR simulation method. Like~\cite{wang_surface_2015} that applied two-step simulations for transmon qubits, we do simulation for the full 3D fluxonium qubit in a commercial high-frequency electromagnetic solver (Ansys HFSS), aiming to solve electric field distribution far from Josephson junction parts, and use low-frequency electromagnetic simulation software (Ansys Maxwell) to extract electric field information close to Josephson junctions. Here we give more detailed information for metal-substrate (MS) interface SPR and note total SPR for fluxonium substrate-air (SA) interface is roughly $3.6\times 10^{-4}$ excluding regions close to small junction. Metal-air (MA) SPR is small compared to MS. Except MS, we further take dielectric layer inside Josephson junctions into account.



\begin{figure} [b]
    \centering
    \includegraphics[width=0.8\linewidth]{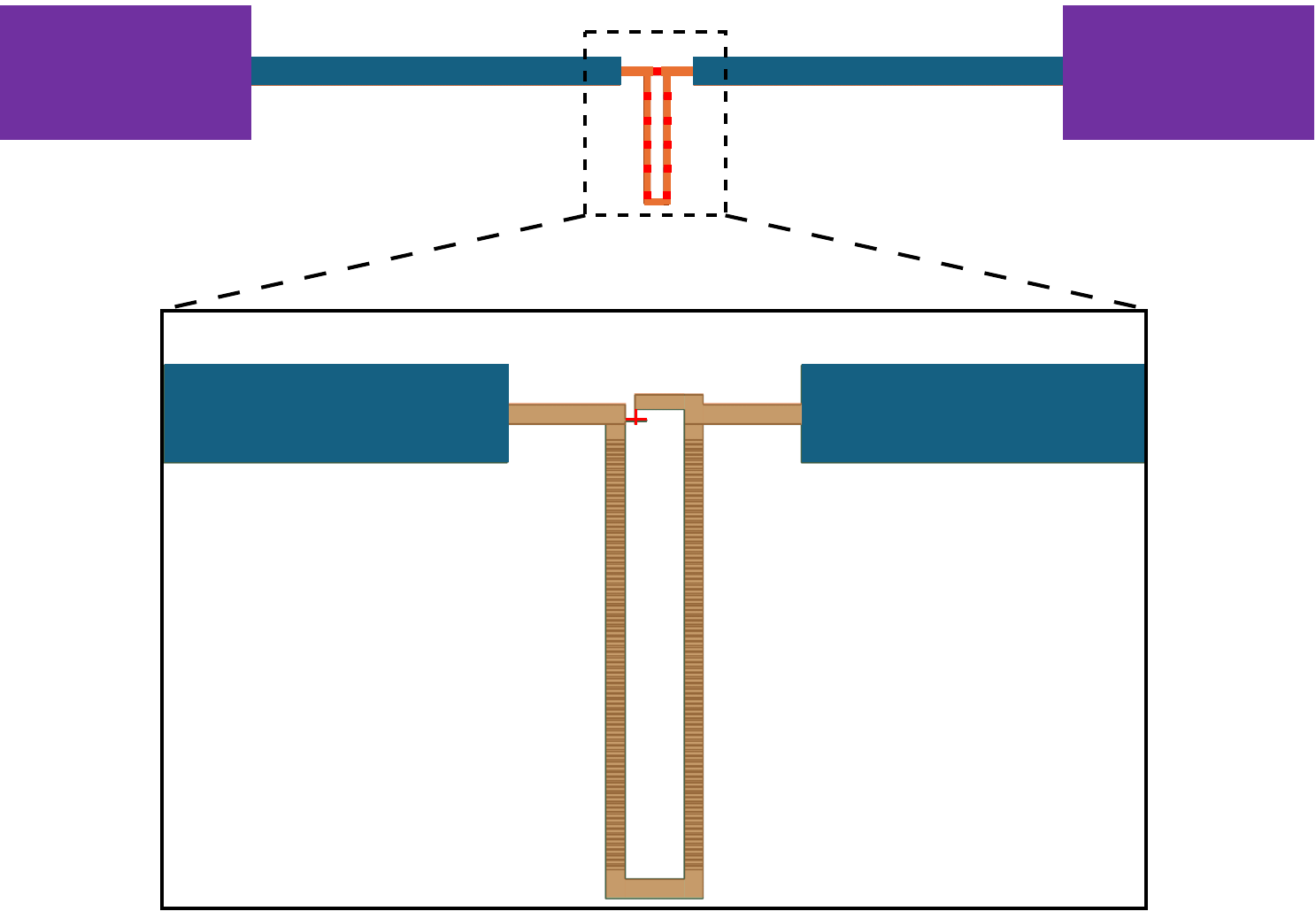}
    \caption{\textbf{Fluxonium qubit in global and local simulations}}
    \label{Fluxonium_in_simulation}
\end{figure}

Fig. \ref{Fluxonium_in_simulation} and its subfigure show the qubit part in global and local simulation, respectively. In the global simulation, we display the whole fluxonium qubit but only take information for large leads (in blue) and pads (in purple). The small junction is represented by one lumped element, and the junction chain is in the form of ten lumped elements, each of which stands for 16.6 large junctions in the chain with integrated capacitance and inductance value. Applying multiple lumped elements to represent junction chain is to better imitate junction chain's electric field in global simulation. In local simulation we draw out details of junction chain and small junction, and focus on obtaining electric field distribution for junction array (in brown) and the small junction (in red). 
We assume a 1.75 nm barrier thickness for all Josephson junction in this fluxonium qubit (transmon's barrier thickness is assumed to be 1.5 nm due to much larger critical current density $J_c$). Electric field results in two simulations are related to each other by normalizing electric field energy inside the small junction, i.e. the transfer coefficient factor: $$F^2 = \frac{U_{j,HFSS}}{U_{j, Maxwell}} = \frac{C_{l}V_l^2}{2\int_{V}UdV} = \frac{ C_l \left( \int_{0}^{L} E_{//} \, dl \right)^2}{\int_{V} \varepsilon_0 \varepsilon_r E^2 \, dx \, dy \, dz},$$ $E_{//}$ is parallel electric field along lumped element's middle line and $C_l$ is lumped element capacitance. Denominator includes energy volume integral to the small junction dielectric layer.

\begin{table}
    \centering
    \renewcommand{\arraystretch}{1}
    \small
    \begin{tabular}{|c|c|c|c|c|c|c|}
        \hline
        & MS pads & MS leads & MS leads & MS leads & Small junction & Junction chains\\
        & & far from junctions & under junction chain & near small junction(s) & & \\
         \hline
        Fluxonium & 0.78 & 2.1 & 1.3 & 18.7 & 1160 & 200 
        \\ \hline
        Transmon & 0.79 & 0.15 & / & 4.3 & 651 & / \\
         \hline 
    \end{tabular}
    \caption{\textbf{Total SPR in parts for fluxonium and tunable transmon device.} All participation ratio results need to further multiplied by $10^{-4}$. For the fluxonium device, as shown in \ref{Fluxonium_in_simulation}, 'MS pads' simply refer to purple parts, and 'MS leads near small junction' include the red cross pattern for small junction. Dielectric layer right under the junction chain falls into the category of 'MS leads under junction chain'. The rest of leads, including dark blue parts and parts of brown area, are classified into 'MS leads far from junctions'. As for transmon, dividing point between near and far from junctions is 1 $\mu$m to two Josephson junctions.}
    \label{SPR_table}
\end{table}

In table \ref{SPR_table} we show a tunable transmon \cite{liu_observation_2024} and 3D fluxonium device's total SPR numbers for different parts. Note the that first three parts contains large volume and dielectric loss caused by TLS inside these area can be modeled using a uniform loss tangent value, which leads to a background decay rate. The rest three parts, on the other hand, have enormous participation ratio. But TLS in these area may either not exist in the frequency range of our spectroscopy by luck, or it can show up as discrete peak feature.

\subsection{TLS dipole moment estimation}
According to our experience to discrete TLS analysis, qubit-TLS coupling $g/(2\pi)$ ranges from 20 kHz to 80 kHz, and average value is around 55 kHz. The electric field inside small junction of fluxonium is derived as $\widehat{E}=\frac{\widehat{Q}}{Cd}=\frac{4E_c\widehat{n}}{e_0d}$, where $e_0$ is elementary charge and d is junction oxide layer thickness (1.75 nm as assumed before). Therefore, the oscillation field strength $\lvert E\rvert$ = 630 V/m. An crude estimation of average electric field strength in the junction chain is $\lvert E_0\rvert = \frac{\lvert E\rvert}{n_0}$ = 3.8 V/m, where $n_0$ is the number of junctions in the junction chain. When we consider mutual capacitance between all metal pieces, the result shows that $\lvert E_0\rvert$ ranges from 1.5 V/m to 11 V/m. Using $|E_0|=3.8$ V/m, we estimate average effective TLS dipole of $p_z\cos\theta=2\hbar g/E_0=1.2$ eÅ, where $p_z$ is the electric dipole moment along the direction of the field, and $\theta=\arctan\frac{\epsilon}{\Delta}$ is the TLS mixing angle between the asymmetry energy and tunneling energy.


\end{document}